\documentclass[english,prd,english,notitlepage,nofootinbib]{revtex4}
\usepackage[latin9]{inputenc}
\usepackage{babel}
\usepackage{amsmath}
\usepackage{amssymb}
\usepackage{graphicx}
\usepackage{esint}
\usepackage[unicode=true,pdfusetitle,
 bookmarks=true,bookmarksnumbered=false,bookmarksopen=false,
 breaklinks=false,pdfborder={0 0 1},backref=false,colorlinks=false]
 {hyperref}

\makeatletter

\providecommand{\tabularnewline}{\\}
\newcommand{\lyxdot}{.}

\@ifundefined{textcolor}{}
{%
 \definecolor{BLACK}{gray}{0}
 \definecolor{WHITE}{gray}{1}
 \definecolor{RED}{rgb}{1,0,0}
 \definecolor{GREEN}{rgb}{0,1,0}
 \definecolor{BLUE}{rgb}{0,0,1}
 \definecolor{CYAN}{cmyk}{1,0,0,0}
 \definecolor{MAGENTA}{cmyk}{0,1,0,0}
 \definecolor{YELLOW}{cmyk}{0,0,1,0}
}

\usepackage{babel}

\providecommand{\tabularnewline}{\\}

\@ifundefined{textcolor}{}{%
 \definecolor{BLACK}{gray}{0}
 \definecolor{WHITE}{gray}{1}
 \definecolor{RED}{rgb}{1,0,0}
 \definecolor{GREEN}{rgb}{0,1,0}
 \definecolor{BLUE}{rgb}{0,0,1}
 \definecolor{CYAN}{cmyk}{1,0,0,0}
 \definecolor{MAGENTA}{cmyk}{0,1,0,0}
 \definecolor{YELLOW}{cmyk}{0,0,1,0}
}

\usepackage{babel}

\makeatother

\begin{document}

\title{GPDs from charged current meson production in $ep$ experiments}

\author{Marat~Siddikov, Iv\'an~Schmidt}

\address{Departamento de F\'isica, Universidad T\'ecnica Federico Santa Mar\'ia,\\
 y Centro Cient\'ifico - Tecnol\'ogico de Valpara\'iso, Casilla 110-V, Valpara\'iso,
Chile}
\begin{abstract}
We suggest that generalized parton distributions can be probed in
charged current meson production process, $ep\to\nu_{e}\pi^{-}p$.
In contrast to pion photoproduction, this process is sensitive to
the unpolarized GPDs $H,\,E$, and for this reason has a very small
contamination by higher twist and Bethe-Heitler type contributions.
Since all produced hadrons are charged, we expect that the kinematics
of this process could be easily reconstructed. We estimated the cross-sections
in the kinematics of upgraded 12 GeV Jefferson Laboratory experiments
and found that thanks to large luminosity the process can be measured
with reasonable statistics. 
\end{abstract}


\maketitle

\section{Introduction}

Understanding the structure of the hadrons presents a challenging
problem both from the theoretical and experimental viewpoints. This
structure is parametrized nowadays in terms of the so-called generalized parton
distributions (GPDs), which can be studied 
in a wide class of processes~\cite{Ji:1998xh,Collins:1998be}. The
early analysis were mostly based on experimental data on deeply virtual
Compton scattering (DVCS)~\cite{Dupre:2017hfs} and deeply virtual
meson production (DVMP)~\cite{Ji:1998xh,Collins:1998be,Mueller:1998fv,Ji:1996nm,Ji:1998pc,Radyushkin:1996nd,Radyushkin:1997ki,Radyushkin:2000uy,Collins:1996fb,Brodsky:1994kf,Goeke:2001tz,Diehl:2000xz,Belitsky:2001ns,Diehl:2003ny,Belitsky:2005qn,Kubarovsky:2011zz},
although it was soon realized that in view of the rich structure of GPDs, as
well as from certain complications with GPD extraction from pion electroproduction~\cite{Kubarovsky:2011zz,Ahmad:2008hp,Goloskokov:2009ia,Goloskokov:2011rd,Goldstein:2012az},
additional channels were needed. It was then suggested that GPDs might be
accessed in $\rho$-meson photoproduction~\cite{Anikin:2009bf,Diehl:1998pd,Mankiewicz:1998kg,Mankiewicz:1999tt,Boussarie:2017umz},
timelike Compton Scattering~\cite{Berger:2001xd,Pire:2008ea,Boer:2015fwa},
exclusive pion- or photon-induced lepton pair production~\cite{Muller:2012yq,Sawada:2016mao},
and heavy charmonia photoproduction~\cite{Ivanov:2004vd,Ivanov:2015hca}
(gluon GPDs). The forthcoming results from the upgraded JLAB~\cite{Kubarovsky:2011zz},
COMPASS~\cite{Gautheron:2010wva,Kouznetsov:2016vvo,Ferrero:2012ega,Sandacz:2016kwh,Sandacz:2017ctv,Silva:2013dta}
as well as from J-PARC~\cite{Sawada:2016mao,Kroll:2016kvd}, hopefully
will enrich and enhance the early data from HERA and 6 GeV JLab experiments,
as well as improve our understanding of the proton GPDs.

Recently we suggested that GPDs could be studied in neutrino-induced
deeply virtual meson production ($\nu$DVMP)~\cite{Kopeliovich:2012dr}
of the pseudoscalar mesons ($\pi,\,K,\,\eta$), using the high-intensity
\textsc{NuMI} beam at Fermilab~\cite{Drakoulakos:2004gn}. The main
advantage of this process is that contamination by twist-3 effects~\cite{Kopeliovich:2014pea}
is small, which implies that GPDs could be accessed at moderate virtualities
$Q^{2}$, provided that the next-to-leading order (NLO) corrections are
included~\cite{Siddikov:2016zmt}. In the Bjorken limit, neglecting the
masses of pions and kaons, we may get information about a full flavor
structure of GPDs. A suppression of Cabibbo forbidden, strangeness
changing processes can be avoided if kaon production is accompanied
by the conversion of a nucleon to strange baryons $\Lambda$ and $\Sigma^{\pm,0}$;
in such processes the transition GPDs are related by $SU(3)$ relations
\cite{Frankfurt:1999fp} to linear combinations of different flavor
components of the nucleon GPDs. Recently it was suggested in~\cite{Pire:2015iza,Pire:2015vxa,Pire:2016jtr,Pire:2017lfj,Pire:2017tvv}
that this approach could be extended to $D$-meson production, a challenge
for future high-energy neutrino experiments. 

In this paper we extend our previous studies to the case of charged
current meson (pion) production in electron-induced processes, such as
\emph{e.g}. $ep\to\nu_{e}\pi^{-}p$. The feasibility to study charged
currents in JLAB kinematics has been demonstrated earlier in~\cite{Androic:2013rhu}.
It is expected that after upgrade even higher luminosities up to $\mathcal{L}=10^{38}{\rm cm}^{-2}\cdot s^{-1}$
will be achieved~\cite{Alcorn:2004sb}, which implies that the DVMP
cross-section could be measured with reasonable statistics. Since
all produced hadrons are charged, the reconstruction of the kinematics
of the process, despite of undetectability of neutrinos, should not
present major difficulties. As will be shown below, the cross-section
of this process on unpolarized targets is mostly sensitive to the GPDs $H_{u},\,H_{d}$,
providing important constraints on available parametrizations, as
well as testing the GPD universality. Similar to the case of neutrino-production,
this process has smaller contamination by higher twist effects compared
to DVMP. 

For the sake of brevity and conciseness, in this paper we do not consider
other processes, where flavor multiplet partners of pions and protons
are produced and which could be used to test other flavor combinations
of GPDs~\cite{Kopeliovich:2012dr}. 

The paper is organized as follows. In Section~\ref{sec:DVMP_Xsec}
we describe the framework used for the evaluation of pion production,
taking into account NLO corrections. In Sections~\ref{sec:BH} and~\ref{sec:Tw3}
we review the contaminating corrections due to Bethe-Heitler mechanism
and twist-three contributions, due to poorly known transversity GPDs.
Finally, in Section~\ref{sec:Results} we present numerical results
and make conclusions.

\section{Cross-section of the $\nu$DVMP process}

\label{sec:DVMP_Xsec}
\begin{figure}[htp]
\includegraphics[height=3cm]{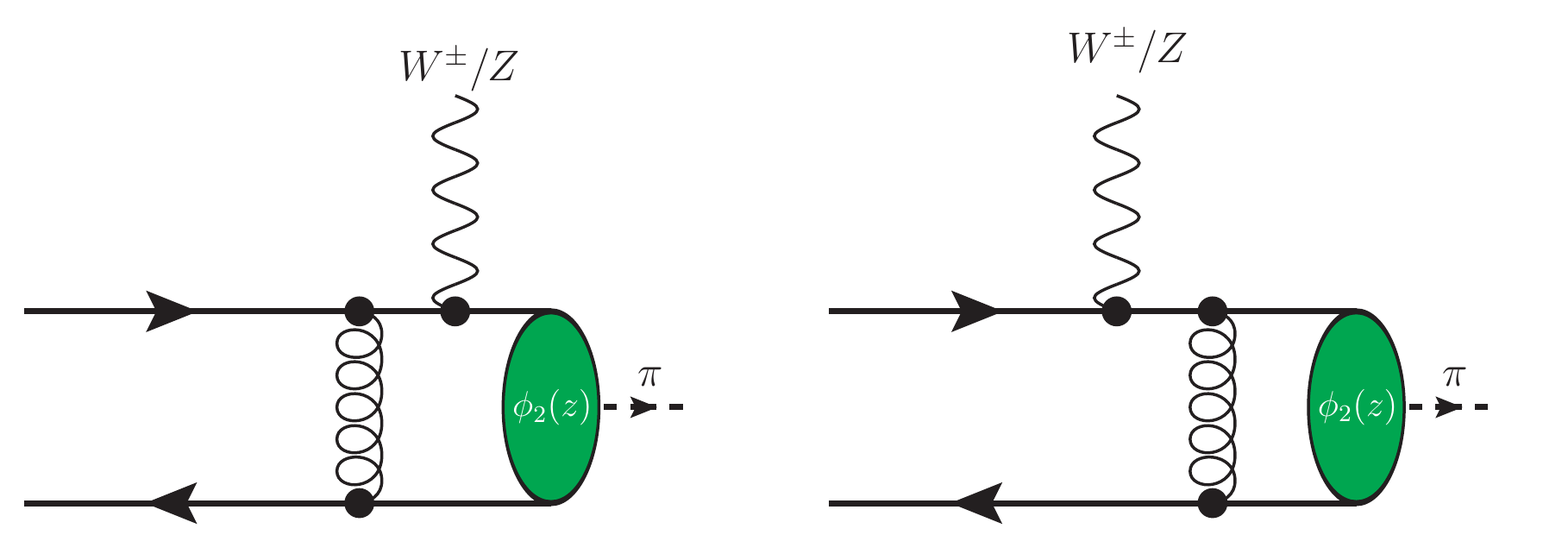} \includegraphics[height=3cm]{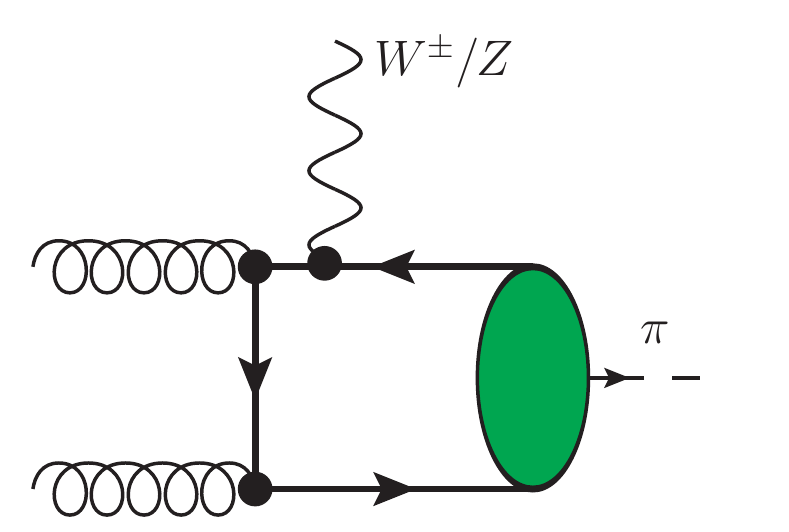}\protect\caption{\label{fig:DVMPLO}Leading-order contributions to the DVMP hard coefficient
functions. Green blob stands for the pion wave function. Additional
diagrams (not shown) may be obtained reversing the directions of the quark
lines and in the case of the last diagram, permuting also the vector boson
vertices.}
\end{figure}

\begin{figure}[htp]
\includegraphics[height=3cm]{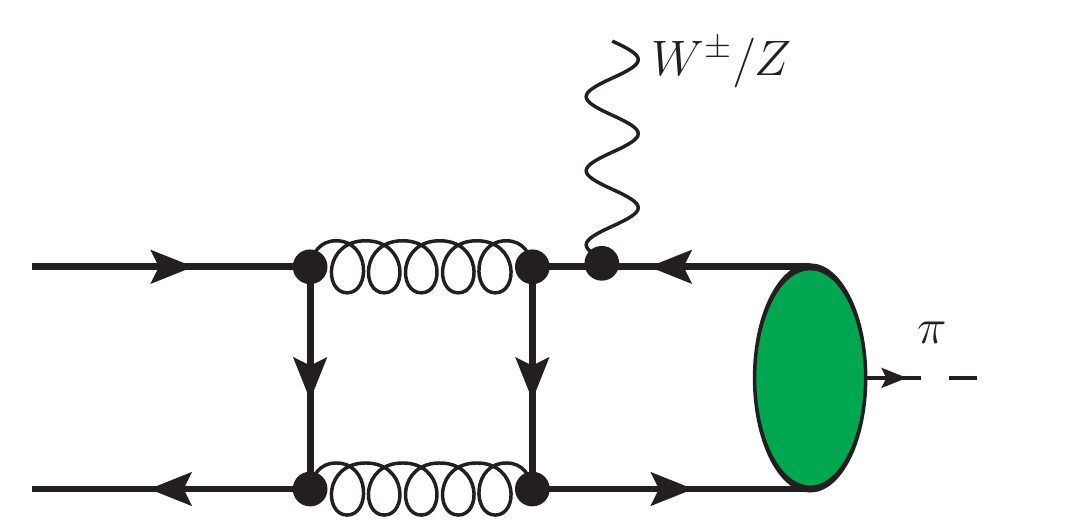}\includegraphics[height=3cm]{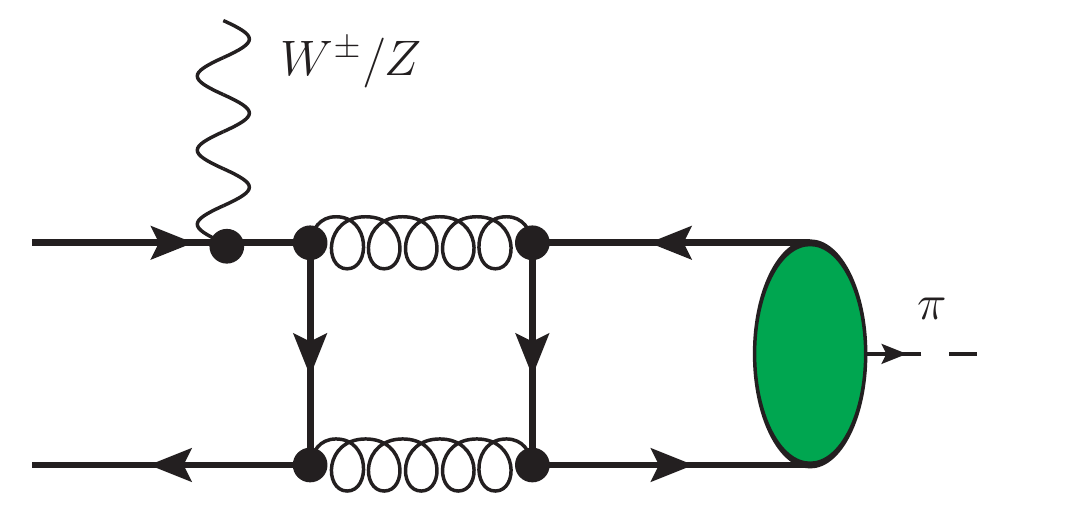}\protect\caption{\label{fig:DVMPNLO-1}Sea quark contributions to the DVMP, which appear
in the next-to-leading-order contributions. Additional diagrams (not
shown) may be obtained by reversing the directions of the quark lines.}
\end{figure}

The cross-section of pion production in charged
current DVMP has a form 
\begin{align}
\frac{d\sigma}{dt\,dx_{B}dQ^{2}} & =\Gamma\sum_{\nu\nu'}\mathcal{A}_{\nu',\nu L}^{*}\mathcal{A}_{\nu',\nu L},\label{eq:sigma_def}
\end{align}
where $t=\left(p_{2}-p_{1}\right)^{2}$ is the momentum transfer to
the proton, $Q^{2}=-q^{2}$ is the virtuality of the charged boson,
$x_{B}=Q^{2}/(2p\cdot q)$ is the Bjorken variable, the subscript
indices $\nu$ and $\nu'$ in the amplitude $\mathcal{A}$ refer to
helicity states of the baryon before and after interaction, and the
letter $L$ reflects the fact that in the Bjorken limit the dominant
contribution comes from the longitudinally polarized massive bosons
$W^{\pm}$~\cite{Ji:1998xh,Collins:1998be}. The kinematic factor
$\Gamma$ included in equation~(\ref{eq:sigma_def}) is given
explicitly,  for the charged current case,  by 
\begin{align}
\Gamma & =\frac{G_{F}^{2}f_{M}^{2}x_{B}^{2}\left(1-y-\frac{\gamma^{2}y^{2}}{4}\right)}{64\pi^{3}Q^{2}\left(1+Q^{2}/M_{W}^{2}\right)^{2}\left(1+\gamma^{2}\right)^{3/2}},
\end{align}
where $\theta_{W}$ is the Weinberg angle, $M_{W}$ is the mass of
the heavy bosons $W^{\pm}$, $G_{F}$ is the Fermi constant, $f_{\pi}$
is the pion decay constant, and we used the shorthand notations 
\begin{equation}
\gamma=\frac{2\,m_{N}x_{B}}{Q},\quad y=\frac{Q^{2}}{s_{ep}\,x_{B}}=\frac{Q^{2}}{2m_{N}E_{e}\,x_{B}}.\label{eq:elasticity}
\end{equation}
where $E_{e}$ is the electron energy in the target rest frame. In
Bjorken kinematics, the amplitude~$\mathcal{A}_{\nu',\nu L}$ factorizes
into a convolution of hard and soft parts, 
\begin{equation}
\mathcal{A}_{\nu',\nu}=\int_{-1}^{+1}dx\sum_{q=u,d,s,g}\,\sum_{\lambda\lambda'}\mathcal{H}_{\nu'\lambda',\nu\lambda}^{q}\mathcal{C}_{\lambda\lambda'}^{q},\label{eq:M_conv}
\end{equation}
where $x$ is the average light-cone fraction of the parton, the superscript
$q$ is its flavor, $\lambda$ and $\lambda'$ are the helicities
of the initial and final partons, and $\mathcal{C}_{\lambda'\nu',\lambda\nu}^{q}$
is the hard coefficient function, which will be specified later. The
soft matrix element $\mathcal{H}_{\nu'\lambda',\nu\lambda}^{q}$ in~(\ref{eq:M_conv_2})
is diagonal in quark helicities ($\lambda,\,\lambda'$), and for the
twist-2 GPDs has the form, 
\begin{align}
\mathcal{H}_{\nu'\lambda',\nu\lambda}^{q} & =\frac{2\delta_{\lambda\lambda'}}{\sqrt{1-\xi^{2}}}\left(-g_{A}^{q}\left(\begin{array}{cc}
\left(1-\xi^{2}\right)H^{q}-\xi^{2}E^{q} & \frac{\left(\Delta_{1}+i\Delta_{2}\right)E^{q}}{2m}\\
-\frac{\left(\Delta_{1}-i\Delta_{2}\right)E^{q}}{2m} & \left(1-\xi^{2}\right)H^{q}-\xi^{2}E^{q}
\end{array}\right)_{\nu'\nu}\right.\label{eq:HAmp}\\
 & +\left.{\rm sgn}(\lambda)g_{V}^{q}\left(\begin{array}{cc}
-\left(1-\xi^{2}\right)\tilde{H}^{q}+\xi^{2}\tilde{E}^{q} & \frac{\left(\Delta_{1}+i\Delta_{2}\right)\xi\tilde{E}^{q}}{2m}\\
\frac{\left(\Delta_{1}-i\Delta_{2}\right)\xi\tilde{E}^{q}}{2m} & \left(1-\xi^{2}\right)\tilde{H}^{q}-\xi^{2}\tilde{E}^{q}
\end{array}\right)_{\nu'\nu}\right),\nonumber 
\end{align}
where the constants $g_{V}^{q},\,g_{A}^{q}$ are the vector and axial
current couplings to quarks, and the leading twist GPDs $H^{q},\,E^{q},\,\tilde{H}^{q}$
and $\tilde{E}^{q}$ are functions of the variables $\left(x,\,\xi,\,t,\,\mu_{F}^{2}\right)$,
where the skewness $\xi$ is related to the light-cone momenta of protons
$p_{1,2}$ as $\xi=\left(p_{1}^{+}-p_{2}^{+}\right)/\left(p_{1}^{+}+p_{2}^{+}\right)$,
the invariant momentum transfer is $t=\Delta^{2}=\left(p_{2}-p_{1}\right)^{2}$,
and $\mu_{F}$ is the factorization scale (see e.g.~\cite{Goeke:2001tz,Diehl:2003ny}
for details of the kinematics). For the processes in which the baryonic
state changes, e.g. $ep\to\nu_{e}\pi_{0}n$, the transition GPDs can
be linearly related via $SU(3)$ relations~\cite{Frankfurt:1999fp}
to ordinary GPDs. For this reason, (\ref{eq:M_conv}) may be effectively
rewritten as 
\begin{equation}
\mathcal{A}_{\nu',\nu}=\int_{-1}^{+1}dx\sum_{q=u,d,s}\,\sum_{\lambda}\mathcal{H}_{\nu'\lambda,\nu\lambda}^{q}\mathcal{C}_{\lambda}^{q}.\label{eq:M_conv_2}
\end{equation}
where $\mathcal{C}_{\lambda}^{q}$ is the diagonal term of the helicity
matrix in the hard coefficient function. Its evaluation is quite straightforward,
and in leading order over $\alpha_{s}$ it gets contributions
from the diagrams shown schematically in Figure~\ref{fig:DVMPLO}.
In fact, it has been studied both for pion electroproduction~\cite{Vanderhaeghen:1998uc,Mankiewicz:1998kg,Goloskokov:2006hr,Goloskokov:2007nt,Goloskokov:2008ib,Goloskokov:2011rd,Goldstein:2012az}
and neutrinoproduction~\cite{Kopeliovich:2012dr}. For the processes
in which the baryon does not change its internal state, there are additional
contributions from gluon GPDs, as shown in the rightmost pane of Figure~\ref{fig:DVMPLO}.
These corrections are small for JLAB kinematics, yet give a 
contributions at higher energies. In next-to-leading order, the
coefficient function includes an additional gluon attached in all
possible ways to all diagrams in Figure~\ref{fig:DVMPLO}, as well
as additional contributions from sea quarks, as shown in Figure~\ref{fig:DVMPNLO-1}.

Straightforward evaluation of the diagrams shown in the Figures~\ref{fig:DVMPLO},\ref{fig:DVMPNLO-1}
yields for the coefficient functions
\begin{align}
\mathcal{C}^{q} & =\eta_{-}^{q}c_{-}^{(q)}\left(x,\,\xi\right)+{\rm sgn}(\lambda)\eta_{+}^{q}c_{+}^{(q)}\left(x,\,\xi\right)+\mathcal{O}\left(\frac{m^{2}}{Q^{2}}\right)+\mathcal{O}\left(\alpha_{s}^{2}\left(\mu_{R}^{2}\right)\right)\label{eq:Coef_function}
\end{align}
where the process-dependent flavor factors $\eta_{V\pm}^{q},\,\eta_{A\pm}^{q}$
are given, for the case of pion production, in Table~\ref{tab:DVMP_amps}~\footnote{As was discussed above, for processes with change of internal baryon
structure, we use $SU(3)$ relations~\cite{Frankfurt:1999fp}, which
are valid up to corrections in current quark masses $\sim\mathcal{O}\left(m_{q}\right)$.}.  In~(\ref{eq:Coef_function}) we also introduced the shorthand notation

\begin{table}
\protect\caption{\label{tab:DVMP_amps}The flavor coefficients $\eta_{\pm}^{q}$, for
several pion and kaon production processes discussed in this paper
($q=u,d,s,...$). For the case of CC mediated processes, take $\eta_{V\pm}^{q}=\eta_{\pm}^{q},\quad\eta_{A\pm}^{q}=-\eta_{\pm}^{q}$.}

\global\long\def\arraystretch{1.5}

\begin{tabular}{|c|c|c|c|c|c|c|}
\cline{1-3} \cline{5-7} 
Process  & $\eta_{+}^{q}$  & $\eta_{-}^{q}$  &  & Process  & $\eta_{+}^{q}$  & $\eta_{-}^{q}$\tabularnewline
\cline{1-3} \cline{5-7} 
$e\,p\to\nu_{e}\pi^{-}p$  & $V_{ud}\delta_{qd}$  & $V_{ud}\delta_{qu}$  &  & $e\,p\to\nu_{e}\pi^{0}n$  & $V_{ud}\frac{\delta_{qu}-\delta_{qd}}{\sqrt{2}}$  & $-V_{ud}\frac{\delta_{qu}-\delta_{qd}}{\sqrt{2}}$ \tabularnewline
\cline{1-3} \cline{5-7} 
$e\,p\to\nu_{e}\pi^{0}n$  & $V_{ud}\frac{\delta_{qu}-\delta_{qd}}{\sqrt{2}}$  & $-V_{ud}\frac{\delta_{qu}-\delta_{qd}}{\sqrt{2}}$  &  & $e\,n\to\nu_{e}\pi^{-}n$  & $V_{ud}\delta_{qu}$  & $V_{ud}\delta_{qd}$\tabularnewline
\cline{1-3} \cline{5-7} 
$e\,p\to\nu_{e}K^{-}p$  & $V_{us}\delta_{qs}$  & $V_{us}\delta_{qs}$  &  & $e\,n\to\nu_{e}K^{0}\Sigma^{-}$  & 0 & $-V_{ud}\left(\delta_{qu}-\delta_{qs}\right)$ \tabularnewline
\cline{1-3} \cline{5-7} 
\end{tabular}
\end{table}

\begin{eqnarray}
c_{\pm}^{(q)}\left(x,\xi\right) & = & \left(\int dz\frac{\phi_{2,\pi}\left(z\right)}{z}\right)\frac{8\pi i}{9}\frac{\alpha_{s}\left(\mu_{R}^{2}\right)f_{M}}{Q}\frac{1}{x\pm\xi\mp i0}\left(1+\frac{\alpha_{s}\left(\mu_{r}^{2}\right)}{2\pi}T^{(1)}\left(\frac{\xi\pm x}{2\xi},\,z\right)\right).\label{eq:c2}
\end{eqnarray}
where $\phi_{2}(z)$ is the twist-2 $\pi-$or $K$-meson distribution
amplitude (DA)~\cite{Kopeliovich:2011rv}. The function $T^{(1)}\left(v,\,z\right)$
in~(\ref{eq:c2}) encodes NLO corrections to the coefficient function.
As was explained in~\cite{Belitsky:2001nq,Ivanov:2004zv,Diehl:2007hd},
it is related by analytical continuation to the loop correction to
$\bar{q}q$ scattering, and was evaluated and analyzed in detail in
the context of NLO studies of the pion form factor (see~\cite{Braaten:1987yy,Melic:1998qr}
for details and historical discussion). Explicitly, it is given by
\begin{align}
T^{(1)}\left(v,\,z\right) & =\frac{1}{2vz}\left[\frac{4}{3}\left([3+\ln(v\,z)]\,\ln\left(\frac{Q^{2}}{\mu_{F}^{2}}\right)+\frac{1}{2}\ln^{2}\left(v\,z\right)+3\ln(v\,z)-\frac{\ln\bar{v}}{2\bar{v}}-\frac{\ln\bar{z}}{2\bar{z}}-\frac{14}{3}\right)\right.\label{eq:T1}\\
 & +\beta_{0}\left(\frac{5}{3}-\ln(v\,z)-\ln\left(\frac{Q^{2}}{\mu_{R}^{2}}\right)\right)\nonumber \\
 & -\frac{1}{6}\left(2\frac{\bar{v}\,v^{2}+\bar{z}\,z^{2}}{(v-z)^{3}}\left[{\rm Li}_{2}(\bar{z})-{\rm Li}_{2}(\bar{v})+{\rm Li}_{2}(v)-{\rm Li}_{2}(z)+\ln\bar{v}\,\ln z-\ln\bar{z}\,\ln v\right]\right.\nonumber \\
 & +2\frac{v+z-2v\,z}{(v-z)^{2}}\ln\left(\bar{v}\bar{z}\right)+2\left[{\rm Li}_{2}(\bar{z})+{\rm Li}_{2}(\bar{v})-{\rm Li}_{2}(z)-{\rm Li}_{2}(v)+\ln\bar{v}\,\ln z+\ln\bar{z}\,\ln v\right]\nonumber \\
 & +\left.\left.4\frac{v\,z\,\ln(v\,z)}{(v-z)^{2}}-4\ln\bar{v}\,\ln\bar{z}-\frac{20}{3}\right)\right],\nonumber 
\end{align}
where $\beta_{0}=\frac{11}{3}N_{c}-\frac{2}{3}N_{f}$, ${\rm Li}_{2}(z)$
is the dilogarithm function, and $\mu_{R}$ and $\mu_{F}$ are the
renormalization and factorization scales respectively. For processes
when the internal state of the hadron is not changed, additional contributions
come from the gluons and singlet (sea) quarks~\cite{Belitsky:2001nq,Ivanov:2004zv,Diehl:2007hd}~\footnote{For the sake of simplicity, we follow~\cite{Diehl:2007hd} and assume
that the factorization scale $\mu_{F}$ is the same for both the generalized
parton distribution and the pion distribution amplitude.}, 
\begin{eqnarray}
c^{(g)}\left(x,\xi\right) & = & \left(\int dz\frac{\phi_{2,\pi}\left(z\right)}{z\,(1-z)}\right)\frac{2\pi i}{3}\frac{\alpha_{s}\left(\mu_{R}^{2}\right)f_{M}}{Q}\frac{\xi}{\left(\xi+x-i0\right)\left(\xi-x-i0\right)}\left(1+\frac{\alpha_{s}\left(\mu_{r}^{2}\right)}{4\pi}\mathcal{I}^{(g)}\left(\frac{\xi-x}{2\xi},\,z\right)\right),\label{eq:c2-2}\\
\mathcal{I}^{(g)}\left(v,\,z\right) & = & \left(\ln\left(\frac{Q^{2}}{\mu_{F}^{2}}\right)-1\right)\left[\frac{\beta_{0}}{2}+C_{A}\left[\left(1-v\right)^{2}+v^{2}\right]\left(\frac{\ln\left(1-v\right)}{v}+\frac{\ln v}{1-v}\right)-\frac{C_{F}}{2}\left(\frac{v\ln v}{1-v}+\frac{\left(1-v\right)\ln\left(1-v\right)}{v}\right)\right.\nonumber \\
 &  & +\left.C_{F}\left(\frac{3}{2}+2\,z\ln\left(1-z\right)\right)\right]-2\,C_{F}-\frac{\beta_{0}}{2}\left(\ln\left(\frac{Q^{2}}{\mu_{R}^{2}}\right)-1\right)-\frac{C_{F}\left(1-2\,v\right)}{2\,\left(z-v\right)}R\left(z,\,v\right)\nonumber \\
 &  & +\frac{\left(2C_{A}-C_{F}\right)}{4}\left(\frac{v\,\ln^{2}v}{1-v}+\frac{\left(1-v\right)\ln^{2}\left(1-v\right)}{v}\right)+C_{F}(1+3\,z)\ln\left(1-z\right)+\nonumber \\
 &  & +\left(\ln\,v+\ln\left(1-v\right)\right)\left[C_{F}\left(1-z\right)\ln\,z-\frac{1}{4}+2C_{F}-C_{A}\right]\nonumber \\
 &  & +\frac{C_{A}}{2}\left(\ln\left(z\,(1-z)\right)-2\right)\left[\frac{v\,\ln v}{1-v}+\frac{\left(1-v\right)\,\ln\left(1-v\right)}{v}\right]\label{eq:T1-a}\\
 &  & +C_{F}z\,\ln^{2}\left(1-z\right)+\frac{C_{A}}{2}\left(1-2\,v\right)\ln\left(\frac{v}{1-v}\right)\left[\frac{3}{2}+\ln\left(z\,\left(1-z\right)\right)+\ln\left(v\,\left(1-v\right)\right)\right]\nonumber \\
 &  & +\left(C_{F}\left((z-v)^{2}-v\,(1-v)\right)-\left(C_{F}-\frac{C_{A}}{2}\right)(z-v)(1-2v)\right)\times\nonumber \\
 &  & \times\left[-\frac{R(z,v)}{(z-v)^{2}}+\frac{\ln v+\ln z-\ln\left(1-v\right)-\ln\left(1-z\right)}{2\left(z-v\right)}+\frac{(z-v)^{2}-v(1-v)}{(z-v)^{3}}H(z,v)\right]\nonumber \\
 &  & +\left\{ \frac{}{}z\to1-z\right\} ,\nonumber \\
C_{F} & = & \frac{N_{c}^{2}-1}{2N_{c}},\quad C_{A}=N_{c}.
\end{eqnarray}
\begin{eqnarray}
c_{\pm}^{(s)}\left(x,\xi\right) & = & -\left(\int dz\frac{\phi_{2,\pi}\left(z\right)}{z\,(1-z)}\right)\frac{4i\alpha_{s}^{2}\left(\mu_{R}^{2}\right)f_{M}}{9\,Q}\mathcal{I}^{(s)}\left(\frac{x\pm\xi}{2\xi},\,z\right),\label{eq:c2-1}\\
\mathcal{I}^{(s)}\left(v,\,z\right) & = & \left(1-2\,v\right)\left(\frac{\ln v}{1-v}+\frac{\ln(1-v)}{v}\right)\ln\left(\frac{Q^{2}z}{\mu_{F}^{2}}\right)+\frac{1-2v}{2}\left[\frac{\ln^{2}v}{1-v}+\frac{\ln^{2}\left(1-v\right)}{v}\right]\label{eq:T1-b}\\
 &  & -\frac{R(v,\,z)}{z-v}-\frac{\left(1-v\right)\ln\left(1-v\right)-v\ln v}{v\left(1-v\right)}+\frac{\left(z-v\right)^{2}-v\left(1-v\right)}{\left(z-v\right)^{2}}H\left(v,z\right)+\left\{ \frac{}{}z\to1-z\right\} ,\nonumber \\
R\left(v,\,z\right) & = & z\ln v+(1-z)\,\ln\left(1-v\right)+z\,\ln z+\left(1-v\right)\ln\left(1-v\right),\\
H\left(v,\,z\right) & = & {\rm Li}_{2}\left(1-v\right)-{\rm Li}_{2}\left(v\right)+{\rm Li}_{2}\left(z\right)-{\rm Li}_{2}\left(1-z\right)+\ln v\,\ln\left(1-z\right)-\ln\left(1-v\right)\ln\,z.
\end{eqnarray}
 Some coefficient functions have non-analytic behavior $\sim\ln^{2}v$
for small $v\approx0$ ($x=\pm\xi\mp i0$), which signals that a collinear
approximation might be not valid near this point. This singularity
in the collinear limit occurs due to the omission of the small transverse
momentum $l_{M,\perp}$ of the quark inside a meson~\cite{Goloskokov:2009ia},
and for this reason the contribution of the region $|v|\sim l_{M,\perp}^{2}/Q^{2}$
should be treated with due care for finite $Q^{2}$ (beyond the Bjorken limit). 
Moreover, a full evaluation of $T^{(1)}\left(v,\,z\right)$
beyond the collinear approximation (taking into account all higher twist
corrections) presents a challenging problem and has not been done
so far. It was observed in~\cite{Diehl:2007hd}, that the singular
terms might be eliminated by a redefinition of the renormalization scale
$\mu_{R}$; however, near the point $v\approx0$ the scale $\mu_{R}^{2}$
becomes soft, $\mu_{R}^{2}\sim z\,v\,Q^{2}\lesssim l_{\perp}^{2}$,
which is another manifestation that nonperturbative effects become
relevant. For this reason, sufficiently large value of $Q^{2}$ should
be used to mitigate contributions of higher twist effects. As we will
see below, for $Q^{2}\approx4$ GeV$^{2}$ the contribution of this
soft region is small, so the collinear factorization is reliable. 

\section{Bethe-Heitler type contribution}

\label{sec:BH}
\begin{figure}[htp]
\includegraphics[scale=0.9]{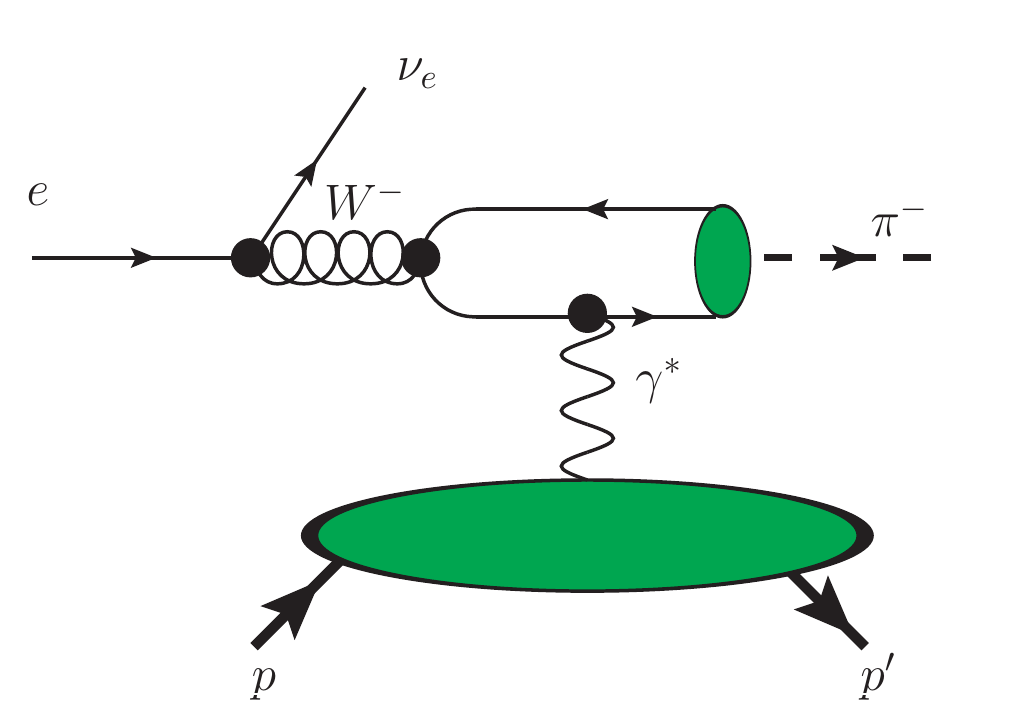}\includegraphics[scale=0.9]{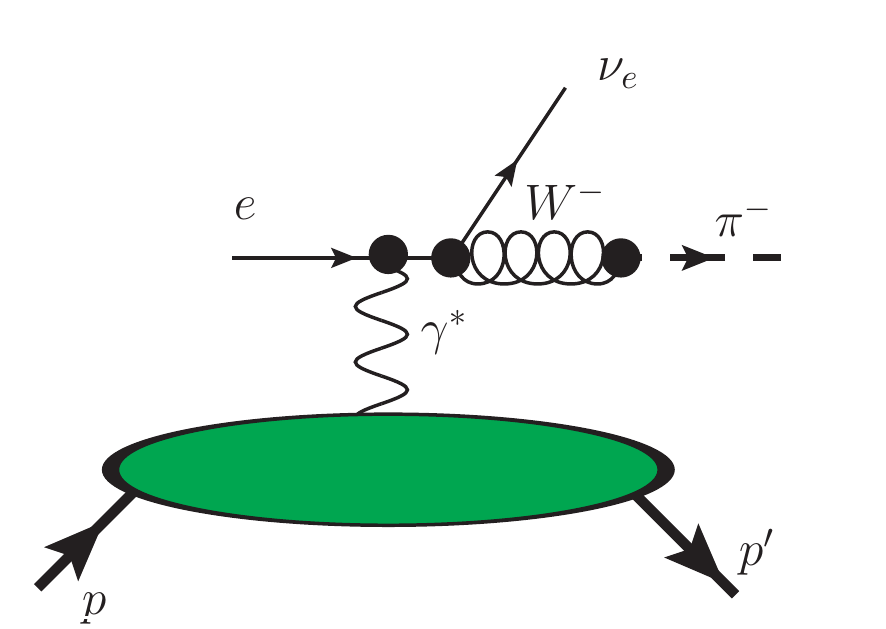}\protect\caption{\label{fig:DVMPBH}Bethe-Heitler contribution in the charged current
DVMP. Formally it is suppressed as $\mathcal{O}\left(\alpha_{{\rm em}}\right)$
compared to DVMP contribution; however, for the asymptotic Bjorken
limit it becomes the dominant mechanism due to relative enhancement by a
kinematical factor $\sim Q^{2}/t\,\alpha_{s}(Q^{2})$. The green blob
stands for the pion wave function.}
\end{figure}

As was found in Ref.~\cite{Kopeliovich:2013ae}, in the
asymptotic Bjorken limit ($Q^{2}\to\infty$) the DVMP contribution
gets overwhelmed by subleading $\mathcal{O}\left(\alpha_{em}\right)$
Bethe-Heitler type (BH) contributions, shown in Figure~\ref{fig:DVMPBH}.
These diagrams have milder suppression at large $Q^{2}$ compared
to DVMP and are additionally enhanced by the $t$-channel photon pole
$\sim1/t$ in the forward kinematics, and for this reason at sufficiently
large $\sim Q^{2}/t$ this mechanism becomes dominant\footnote{As was estimated in~\cite{Kopeliovich:2013ae}, the cross-section
of Bethe-Heitler mechanism becomes comparable to DVMP for $Q^{2}\gtrsim100$~GeV$^{2}$}. 

While for the DVMP amplitude evaluation presented in the previous section the
dominant contribution comes from the longitudinally polarized mediator
boson, for the BH this is no longer true and we have to consider all the
$W$ polarizations. The left diagram in Figure~\ref{fig:DVMPBH}
contains the matrix element

\begin{equation}
\mathcal{A}_{\mu\nu}^{ab}\left(q,\Delta\right)=\frac{1}{f_{\pi}}\int d^{4}x\,e^{-iq\cdot x}\left\langle 0\left|\left(V_{\mu}^{a}(x)-A_{\mu}^{a}(x)\right)J_{\nu}^{em}(0)\right|\pi^{b}\left(q-\Delta\right)\right\rangle ,\label{eq:A_munu_def}
\end{equation}
where $V_{\mu}^{a}(x)$ and $A_{\mu}^{a}(x)$ are the vector and axial
isovector currents. We evaluate the correlator~(\ref{eq:A_munu_def})
perturbatively in the collinear approximation, which is justified
by the intermediate boson large $Q^{2}$ value, and therefore consider only the dominant
contribution from the leading twist-2 pion DA. The evaluation details 
may be found in~\cite{Kopeliovich:2013ae}, while here for the sake of
brevity we will only provide the final result. The cross-section of
the Bethe-Heitler mechanism is given by

\begin{equation}
\frac{d^{4}\sigma^{(BH)}}{dt\,d\ln x_{Bj}\,dQ^{2}d\varphi}=\frac{f_{\pi}^{2}G_{F}^{2}\alpha_{em}^{2}\sum_{n=0}^{2}\mathcal{C}_{n}^{BH}\cos(n\varphi)}{16\,\pi^{2}t^{2}\left(1+\frac{4m_{N}^{2}x_{B}^{2}}{Q^{2}}\right)^{5/2}},\label{eq:XSec_BH}
\end{equation}
where $\varphi$ is the angle between the lepton scattering and the pion production
planes, and in addition to the kinematic variables defined in the
previous section~\ref{sec:DVMP_Xsec} we introduced shorthand notations
\begin{align}
\mathcal{C}_{0}^{BH} & =\mathcal{C}_{2}^{BH}+\frac{m_{N}^{2}}{9\,Q^{2}}\left[4\left(\left(2y^{2}+y-1\right)\left(\phi_{-1}-1\right)x_{B}^{3}\right.\right.\label{eq:BH_0}\\
 & -\left(\left(4\left(\phi_{-1}-1\right)^{2}+\frac{t}{2m_{N}^{2}}\left(4\phi_{-1}^{2}-8\phi_{-1}+5\right)\right)y^{2}-4\left(\left(\phi_{-1}-1\right)^{2}+\frac{t}{4m_{N}^{2}}\left(4\phi_{-1}^{2}-13\phi_{-1}+10\right)\right)y\right.\nonumber \\
 & +\left.\frac{5t}{2m_{N}^{2}}\left(\phi_{-1}-2\right)^{2}+\left(\phi_{-1}-1\right)^{2}\right)x_{B}^{2}+\left(2y-1\right)\frac{t}{m_{N}^{2}}\left(\phi_{-1}-1\right)\left(-\phi_{-1}+y\left(2\phi_{-1}-1\right)+2\right)x_{B}\nonumber \\
 & -\left.4\left(1-2y\right)^{2}\frac{t}{4m_{N}^{2}}\left(\phi_{-1}-1\right)^{2}\right)F_{1}^{2}(t)+2F_{1}(t)F_{2}(t)x_{B}^{2}\left(x_{B}^{2}\left(y+1\right)^{2}-\frac{x_{B}t}{m_{N}^{2}}\left(y+1\right)^{2}\right.\nonumber \\
 & \left.-\frac{t}{m_{N}^{2}}\left(\left(8\phi_{-1}^{2}-24\phi_{-1}+17\right)y^{2}-2\left(8\phi_{-1}^{2}-24\phi_{-1}+19\right)y+10\phi_{-1}^{2}-36\phi_{-1}+35\right)\right)\nonumber \\
 & +F_{2}^{2}\left(\left(y+1\right)^{2}\left(\frac{t}{4m_{N}^{2}}+1\right)x_{B}^{4}-\left(y+1\right)\frac{t}{m_{N}^{2}}\left(\frac{t}{4m_{N}^{2}}-\phi_{-1}+y\left(\frac{t}{4m_{N}^{2}}+2\phi_{-1}-1\right)+2\right)x_{B}^{3}\right.\nonumber \\
 & +\frac{t\,x_{B}^{2}}{m_{N}^{2}}\left(\left(-4\phi_{-1}^{2}+16\phi_{-1}+\frac{t}{4m_{N}^{2}}\left(8\phi_{-1}-7\right)-13\right)y^{2}\right.\nonumber \\
 & \left.+2\left(6\phi_{-1}^{2}-20\phi_{-1}+\frac{t}{4m_{N}^{2}}\left(2\phi_{-1}-1\right)+17\right)y-9\phi_{-1}^{2}+\frac{t}{4m_{N}^{2}}\left(5-4\phi_{-1}\right)+34\phi_{-1}-34\right)\nonumber \\
 & \left.\left.-\left(2y-1\right)\left(\frac{t}{m_{N}^{2}}\right)^{2}\left(\phi_{-1}-1\right)\left(-\phi_{-1}+y\left(2\phi_{-1}-1\right)+2\right)x_{B}+\left(1-2y\right)^{2}\left(\frac{t}{4m_{N}^{2}}\right)^{2}\left(\phi_{-1}-1\right)^{2}\right)\right]\nonumber \\
 & +\mathcal{O}\left(\frac{m_{N}^{4}}{Q^{4}},\frac{t^{2}}{Q^{4}}\right),\nonumber 
\end{align}

\begin{align}
\mathcal{C}_{1}^{BH} & =\frac{K\,m_{N}^{2}}{9Q^{2}}\left[4\left(3\left(-4y+3\left(y-2\right)\phi_{-1}+9\right)x_{B}^{3}\right.\right.\label{eq:BH_1}\\
 & -2\left(\phi_{-1}-1\right)\left(-\left(\frac{5t}{2m_{N}^{2}}+9\right)\phi_{-1}+2y\left(\frac{t\phi_{-1}}{m_{N}^{2}}-\frac{3t}{4m_{N}^{2}}+3\phi_{-1}-6\right)+18\right)x_{B}^{2}\nonumber \\
 & +\left.\frac{3t}{m_{N}^{2}}\left(\phi_{-1}-1\right)\left(-6\phi_{-1}+y\left(4\phi_{-1}-3\right)+3\right)x_{B}-\left(2y-3\right)\frac{6\,t}{m_{N}^{2}}\left(\phi_{-1}-1\right)^{2}\right)F_{1}^{2}(t)\nonumber \\
 & +4F_{1}(t)F_{2}(t)x_{B}^{2}\left(3\left(y-3\right)x_{B}^{2}-12\left(y-3\right)\frac{t}{4m_{N}^{2}}x_{B}-8\frac{t}{4m_{N}^{2}}\left(4y\left(\phi_{-1}-3\right)-5\phi_{-1}+18\right)\left(\phi_{-1}-1\right)\right)\nonumber \\
 & -F_{2}^{2}(t)\left(-6\left(y-3\right)x_{B}^{4}\right.\nonumber \\
 & +\frac{t}{4m_{N}^{2}}\left(-6\left(y-3\right)x_{B}^{2}+12\left(-2y+3\left(y-2\right)\phi_{-1}+3\right)x_{B}+8\left(2y\left(\phi_{-1}-6\right)-\phi_{-1}+18\right)\left(\phi_{-1}-1\right)\right)x_{B}^{2}\nonumber \\
 & \left.\left.+24\left(\frac{t}{4m_{N}^{2}}\right)^{2}\left(x_{B}\left(x_{B}-2\phi_{-1}+2\right)+2\left(\phi_{-1}-1\right)\right)\left(x_{B}\left(y-3\right)-2\left(2y-3\right)\left(\phi_{-1}-1\right)\right)\right)\right]\nonumber \\
 & +\mathcal{O}\left(\frac{m_{N}^{4}}{Q^{4}},\frac{t^{2}}{Q^{4}}\right)\nonumber 
\end{align}
\begin{align}
\mathcal{C}_{2}^{BH} & =-\frac{4\,K^{2}}{9}\left[\left(5x_{B}-4\phi_{-1}+4\right)\left(\phi_{-1}-1\right)F_{1}^{2}(t)+2x_{B}^{2}F_{1}(t)F_{2}(t)\right.\label{eq:BH_2}\\
 & +\left.\left(\left(1+\frac{t}{4m_{N}^{2}}\right)x_{B}^{2}-\frac{5\,t\,x_{B}}{4m_{N}^{2}}\left(\phi_{-1}-1\right)+\frac{t}{m_{N}^{2}}\left(\phi_{-1}-1\right)^{2}\right)F_{2}^{2}(t)\right]\nonumber \\
 & +\mathcal{O}\left(\frac{m_{N}^{2}}{Q^{2}},\,\frac{t}{Q^{2}}\right),\nonumber 
\end{align}

\begin{align}
K^{2} & =\frac{\Delta_{\perp}^{2}}{Q^{2}}\left(1-y-\frac{y^{2}\epsilon^{2}}{4}\right)\\
 & =-\frac{t}{Q^{2}}\left(1-x_{B}\right)\left(1-y-\frac{\epsilon^{2}y^{2}}{4}\right)\left(1-\frac{t_{min}}{t}\right)\left\{ \sqrt{1+\epsilon^{2}}+\frac{4x_{B}\left(1-x_{B}\right)+\epsilon^{2}}{4\left(1-x_{B}\right)}\frac{t-t_{min}}{Q^{2}}\right\} ,\nonumber \\
\epsilon^{2} & =\frac{4m_{N}^{2}x_{B}^{2}}{Q^{2}},\quad t_{min}=-\frac{m_{N}^{2}x_{B}^{2}}{1-x_{B}}+\mathcal{O}\left(\frac{m_{N}^{2}}{Q^{2}},\frac{t}{Q^{2}}\right),\qquad\phi_{-1}=\int_{0}^{1}\,\frac{\phi_{2,\pi}(z)}{z}\,dz.
\end{align}

In the expressions~(\ref{eq:BH_0}-\ref{eq:BH_2}) we use the notation $F_{1,2}(t)$
for the Dirac and Pauli form factors. As we can see, the BH cross-section
is symmetric under the $\varphi\to-\varphi$ transformation. For asymptotically
large $Q^{2}$, the $\mathcal{C}_{1}^{BH}$ harmonic is suppressed
by $\Delta_{\perp}/Q$, whereas $\mathcal{C}_{0}^{BH}\sim\mathcal{C}_{2}^{BH}$,
and therefore the distribution is also symmetric with respect to the $\varphi\to\pi-\varphi$
transformation. 

The interference between the DVMP and BH amplitudes yields an additional
contribution

\begin{align}
\frac{d^{4}\sigma^{(int)}}{dt\,d\ln x_{Bj}\,dQ^{2}d\varphi} & =\frac{f_{\pi}^{2}G_{F}^{2}\,x_{B}\alpha_{em}\alpha_{S}\phi_{-1}\left(\mathcal{C}_{0}^{int}+\mathcal{C}_{1}^{int}\cos\varphi+\mathcal{S}_{1}^{int}\sin\varphi\right)}{36\,\pi^{2}t\,Q^{2}\left(1-\frac{x_{B}}{2}\right)\left(1+\frac{4m_{N}^{2}x_{B}^{2}}{Q^{2}}\right)^{5/2}},\label{eq:XSec_Int}
\end{align}
where 
\begin{align}
\mathcal{C}_{0}^{int} & =-\frac{m_{N}^{2}(1-y)}{Q^{2}}\left(\left(-4\left(1-x_{B}\right)\Re e\mathcal{H}+x_{B}^{2}\,\Re e\mathcal{E}\right)F_{1}+\Re e\mathcal{E}\,F_{2}\frac{t}{4m_{N}^{2}}(x_{B}-2)^{2}+\left(\Re e\mathcal{H}+\Re e\mathcal{E}\right)F_{2}x_{B}^{2}\right)\label{eq:Int_C0}\\
 & \times\left(\left(2\phi_{-1}-3\right)x_{B}^{2}-\frac{t}{m_{N}^{2}}\left(\phi_{-1}-1\right)\left(1+x_{B}\right)\right)+\mathcal{O}\left(\frac{m_{N}^{2}}{Q^{2}},\frac{t}{Q^{2}}\right),\nonumber \\
\mathcal{C}_{1}^{int} & =\frac{K}{3}\,\left[F_{1}\left(2\,\mathcal{R}e\,\mathcal{E}\left(y-3\right)x_{B}^{2}+\mathcal{R}e\,\mathcal{H}\left(4\left(x_{B}-2\right)\left(2y-3\right)\phi_{-1}-4\left(\left(x_{B}-4\right)y+6\right)\right)\right)\right.\label{eq:Int_C1}\\
 & +F_{2}\left(2\,\mathcal{R}e\mathcal{H}\left(y-3\right)x_{B}^{2}+\mathcal{R}e\,\mathcal{E}\left(2x_{B}^{2}\left(y-3\right)-\left(x_{B}-2\right)\frac{t}{4m_{N}^{2}}\left(4\left(2y-3\right)\left(\phi_{-1}-1\right)-2x_{B}\left(y-3\right)\right)\right)\right)\nonumber \\
 & +\left.\mathcal{O}\left(\frac{m_{N}^{2}}{Q^{2}},\frac{t}{Q^{2}}\right)\right]\nonumber \\
\mathcal{S}_{1}^{int} & =\frac{K\left(2-x_{B}\right)}{6}\,\left[F_{1}\left(-2\,\mathcal{I}m\,\mathcal{E}\left(y+1\right)x_{B}^{2}-2\,\mathcal{I}m\,\mathcal{H}\left(x_{B}\left(4\phi_{-1}y-2y-2\phi_{-1}+4\right)-4\left(2y-1\right)\left(\phi_{-1}-1\right)\right)\right)\right.\label{eq:Int_S1}\\
 & +F_{2}\left(-2\,\mathcal{I}m\,\mathcal{H}\left(y+1\right)x_{B}^{2}\right.\nonumber \\
 & \left.-2\,\mathcal{I}m\,\mathcal{E}\left(\left(y+1\right)\left(1+\frac{t}{4m_{N}^{2}}\right)x_{B}^{2}+\frac{t}{2m_{N}^{2}}\left(-2\phi_{-1}y+y+\phi_{-1}-2\right)x_{B}+\left(2y-1\right)\frac{t}{m_{N}^{2}}\left(\phi_{-1}-1\right)\right)\right)\nonumber \\
 & +\left.\mathcal{O}\left(\frac{m_{N}^{2}}{Q^{2}},\frac{t}{Q^{2}}\right)\right],\nonumber 
\end{align}

The angular dependence of the interference term~(\ref{eq:XSec_Int})
has a $\sim\sin\varphi$ term, which stems from the interference of
the vector and axial vector currents in the lepton part of the diagram.
This interference contribution depends only linearly on the target
GPDs and for this reason presents interesting opportunities for studies
at future colliders. 

As we will see below, in JLAB kinematics the contribution of both
BH and interference terms are small, and for this reason it is convenient
to assess their size in terms of the angular harmonics $c_{n},s_{n}$,
normalizing the total cross-section to the cross-section of the dominant
DVMP process as\footnote{Compared to our earlier~\cite{Kopeliovich:2013ae}, we modified the definition
of $c_{0}$ and explicitly took out the unit term in~(\ref{eq:Harmonics}),
in order to have uniform counting $c_{n},\,s_{n}\sim\mathcal{O}\left(\alpha_{{\rm em}}\right)$
for all harmonics. }
\begin{equation}
\frac{d^{4}\sigma^{(tot)}}{dt\,d\ln x_{Bj}\,dQ^{2}d\varphi}=\frac{1}{2\pi}\frac{d^{4}\sigma^{(DVMP)}}{dt\,d\ln x_{Bj}\,dQ^{2}}\left(1+\sum_{n=0}^{2}c_{n}\cos(n\varphi)+s_{1}\sin(\varphi)\right).\label{eq:Harmonics}
\end{equation}

\section{Twist-three corrections}

\label{sec:Tw3}In the Bjorken limit, the dominant contribution comes
from the twist-two GPDs $H,\,E,\,\tilde{H},\,\tilde{E}$. However,
in modern experiments a large part of the data comes from the region
of $Q$ only two or three times larger than the nucleon mass $m_{N}$.
For this reason it is important to assess how large are the omitted
higher-twist contributions. Previously this analysis was done
by us in the context of neutrino-production \cite{Kopeliovich:2014pea},
and here we repeat it for the case of charged current meson production. 

The additional contribution to the amplitude~(\ref{eq:HAmp}) from transversity GPDs
is given by 
\begin{align}
\delta\mathcal{H}_{\nu'\lambda',\nu\lambda}^{q} & =\left(m_{\nu'\nu}^{q}\delta_{\lambda,-}\delta_{\lambda',+}+n_{\nu'\nu}^{q}\delta_{\lambda,+}\delta_{\lambda',-}\right),
\end{align}
where the coefficients $m_{\pm,\pm}^{q}$ and $n_{\pm,\pm}^{q}$ are
linear combinations of the transversity GPDs, 
\begin{align}
m_{--}^{q} & =\frac{\sqrt{-t'}}{4m}\left[2\tilde{H}_{T}^{q}\,+(1+\xi)E_{T}^{q}-(1+\xi)\tilde{E}_{T}^{q}\right],\\
m_{-+}^{q} & =\sqrt{1-\xi^{2}}\frac{t'}{4m^{2}}\tilde{H}_{T}^{q},\\
m_{+-}^{q} & =\sqrt{1-\xi^{2}}\left[H_{T}^{q}-\frac{\xi^{2}}{1-\xi^{2}}E_{T}^{q}+\frac{\xi}{1-\xi^{2}}\tilde{E}_{T}^{q}-\frac{t'}{4m^{2}}\tilde{H}_{T}^{q}\right],\\
m_{++}^{q} & =\frac{\sqrt{-t'}}{4m}\left[2\tilde{H}_{T}^{q}+(1-\xi)E_{T}^{q}+(1-\xi)\tilde{E}_{T}^{q}\right],
\end{align}

\begin{align}
n_{--}^{q} & =-\frac{\sqrt{-t'}}{4m}\left(2\tilde{H}_{T}^{q}+(1-\xi)E_{T}^{q}+(1-\xi)\tilde{E}_{T}^{q}\right),\\
n_{-+}^{q} & =\sqrt{1-\xi^{2}}\left(H_{T}^{q}-\frac{\xi^{2}}{1-\xi^{2}}E_{T}^{q}+\frac{\xi}{1-\xi^{2}}\tilde{E}_{T}^{q}-\frac{t'}{4m^{2}}\tilde{H}_{T}^{q}\right),\\
n_{+-}^{q} & =\sqrt{1-\xi^{2}}\frac{t'}{4m^{2}}\tilde{H}_{T}^{q},\\
n_{++}^{q} & =-\frac{\sqrt{-t'}}{4m}\left(2\tilde{H}_{T}^{q}+(1+\xi)E_{T}^{q}-(1+\xi)\tilde{E}_{T}^{q}\right),
\end{align}
and we introduced a shorthand notation $t'=-\Delta_{\perp}^{2}/(1-\xi^{2})$;
$\Delta_{\perp}=p_{2,\perp}-p_{1,\perp}$ is the transverse part of
the momentum transfer. The coefficient function~(\ref{eq:Coef_function})
gets an additional nondiagonal in parton helicity contribution,

\begin{align}
\delta\mathcal{C}_{\lambda'0,\lambda\mu}^{q} & ==\delta_{\mu,+}\delta_{\lambda,-}\delta_{\lambda',+}\left(S_{A}^{q}-S_{V}^{q}\right)+\delta_{\mu,-}\delta_{\lambda,+}\delta_{\lambda',-}\left(S_{A}^{q}+S_{V}^{q}\right)+\mathcal{O}\left(\frac{m^{2}}{Q^{2}}\right),\label{eq:Coef_function-1}
\end{align}
where we introduced the shorthand notations
\begin{eqnarray}
S_{A}^{q} & = & \int dz\,\left(\left(\eta_{A+}^{q}c_{+}^{(3,p)}\left(x,\xi\right)-\eta_{A-}^{q}c_{-}^{(3,p)}\left(x,\xi\right)\right)+2\left(\eta_{A-}^{q}c_{-}^{(3,\sigma)}\left(x,\xi\right)+\eta_{A+}^{q}c_{+}^{(3,\sigma)}\left(x,\xi\right)\right)\right),\label{eq:SA_def}\\
S_{V}^{q} & = & \int dz\,\left(\left(\eta_{V+}^{q}c_{+}^{(3,p)}\left(x,\xi\right)+\eta_{V-}^{q}c_{-}^{(3,p)}\left(x,\xi\right)\right)+2\left(\eta_{V+}^{q}c_{+}^{(3,\sigma)}\left(x,\xi\right)-\eta_{V-}^{q}c_{-}^{(3,\sigma)}\left(x,\xi\right)\right)\right),\label{eq:SV_def}
\end{eqnarray}
\begin{equation}
c_{+}^{(3,i)}\left(x,\xi\right)=\frac{4\pi i\alpha_{s}f_{\pi}\xi}{9\,Q^{2}}\int_{0}^{1}dz\frac{\phi_{3,i}(z)}{z(x+\xi)^{2}},\quad c_{-}^{(3,i)}\left(x,\xi\right)=\frac{4\pi i\alpha_{s}f_{\pi}\xi}{9\,Q^{2}}\int_{0}^{1}dz\frac{\phi_{3,i}(z)}{(1-z)(x-\xi)^{2}};\label{eq:Tw3_coefFunction}
\end{equation}
and the twist-three pion distributions are defined as 
\begin{eqnarray}
\phi_{3}^{(p)}\left(z\right) & = & \frac{1}{f_{\pi}\sqrt{2}}\frac{m_{u}+m_{d}}{m_{\pi}^{2}}\int\frac{du}{2\pi}e^{i(z-0.5)u}\left\langle 0\left|\bar{\psi}\left(-\frac{u}{2}n\right)\gamma_{5}\psi\left(\frac{u}{2}n\right)\right|\pi(q)\right\rangle ,\label{eq:DA3p}
\end{eqnarray}

\begin{eqnarray}
\phi_{3}^{(\sigma)}\left(z\right) & = & \frac{3i}{\sqrt{2}f_{\pi}}\frac{m_{u}+m_{d}}{m_{\pi}^{2}}\int\frac{du}{2\pi}e^{i(z-0.5)u}\left\langle 0\left|\bar{\psi}\left(-\frac{u}{2}n\right)\sigma_{+-}\gamma_{5}\psi\left(\frac{u}{2}n\right)\right|\pi(q)\right\rangle .\label{eq:DA3s}
\end{eqnarray}

Thanks to symmetry of $\phi_{p}$ and antisymmetry of $\phi_{\sigma}$
with respect to charge conjugation, the dependence on the pion DAs
factorizes in the collinear approximation and contributes only as
the minus one first moment of the linear combination of the twist-3 DAs,
$\phi_{p}(z)+2\phi_{\sigma}(z)$, 
\begin{equation}
\left\langle \phi_{3}^{-1}\right\rangle =\int_{0}^{1}dz\frac{\phi_{3}^{(p)}\left(z\right)+2\phi_{3}^{(\sigma)}\left(z\right)}{z}.
\end{equation}
In the general case the coefficient function ~(\ref{eq:Tw3_coefFunction})
leads to collinear divergencies near the points $x=\pm\xi$, when substituted
to (\ref{eq:M_conv_2}). As was noted in~\cite{Goloskokov:2009ia},
this singularity is naturally regularized by the the small transverse
momentum of the quarks inside the meson. Such regularization modifies~(\ref{eq:Tw3_coefFunction})
to 
\begin{align}
c_{+}^{(3,i)}\left(x,\xi\right) & =\frac{4\pi i\alpha_{s}f_{\pi}\xi}{9\,Q^{2}}\int_{0}^{1}dz\,d^{2}l_{\perp}\frac{\phi_{3,i}\left(z,\,l_{\perp}\right)}{(x+\xi-i0)\left(z(x+\xi)+\frac{2\xi\,l_{\perp}^{2}}{Q^{2}}\right)},\label{eq:c3Plus}\\
c_{-}^{(3,i)}\left(x,\xi\right) & =\frac{4\pi i\alpha_{s}f_{\pi}\xi}{9\,Q^{2}}\int_{0}^{1}dz\,d^{2}l_{\perp}\frac{\phi_{3,i}\left(z,\,l_{\perp}\right)}{(x-\xi+i0)\left((1-z)(x-\xi)-\frac{2\xi\,l_{\perp}^{2}}{Q^{2}}\right)},\label{eq:c3Minus}
\end{align}
where $l_{\perp}$ is the transverse momentum of the quark, and we
tacitly assume absence of any other transverse momenta in the coefficient
function. Due to interference of the leading twist and twist-there
contributions, the total cross-section acquires dependence on the
angle $\varphi$ between lepton scattering and pion production planes,
\begin{align}
\frac{d\sigma}{dt\,dx_{B}dQ^{2}d\varphi} & =\epsilon\frac{d\sigma_{L}}{dt\,dx_{B}dQ^{2}d\varphi}+\frac{d\sigma_{T}}{dt\,dx_{B}dQ^{2}d\varphi}+\sqrt{\epsilon(1+\epsilon)}\cos\varphi\frac{d\sigma_{LT}}{dt\,dx_{B}dQ^{2}d\varphi}\label{eq:sigma_def-1}\\
 & +\epsilon\cos\left(2\varphi\right)\frac{d\sigma_{TT}}{dt\,dx_{B}dQ^{2}d\varphi}+\sqrt{\epsilon(1+\epsilon)}\sin\varphi\frac{d\sigma_{L'T}}{dt\,dx_{B}dQ^{2}d\varphi}+\epsilon\sin\left(2\varphi\right)\frac{d\sigma_{T'T}}{dt\,dx_{B}dQ^{2}d\varphi},\nonumber 
\end{align}
where we introduced the shorthand notations
\[
\epsilon=\frac{1-y-\frac{\gamma^{2}y^{2}}{4}}{1-y+\frac{y^{2}}{2}+\frac{\gamma^{2}y^{2}}{4}}.
\]
\begin{align}
\frac{d\sigma_{L}}{dt\,dx_{B}dQ^{2}d\varphi} & =\frac{\Gamma\,\sigma_{00}}{2\pi\epsilon}\label{eq:sigma_L}\\
\frac{d\sigma_{T}}{dt\,dx_{B}dQ^{2}d\varphi} & =\frac{\Gamma}{2\pi\epsilon}\,\left(\frac{\sigma_{++}+\sigma_{--}}{2}+\frac{1}{2}\sqrt{1-\epsilon^{2}}\frac{\sigma_{++}-\sigma_{--}}{2}\right)\label{eq:sigma_T}\\
\frac{d\sigma_{LT}}{dt\,dx_{B}dQ^{2}d\varphi} & =\frac{\Gamma}{2\pi\epsilon}\,\left({\rm Re}\left(\sigma_{0+}-\sigma_{0-}\right)+\frac{1}{2}\sqrt{\frac{1-\epsilon}{1+\epsilon}}{\rm Re}\left(\sigma_{0+}+\sigma_{0-}\right)\right)\label{eq:sigma_LT}\\
\frac{d\sigma_{TT}}{dt\,dx_{B}dQ^{2}d\varphi} & =-\frac{\Gamma}{2\pi\epsilon}\,{\rm Re}\left(\sigma_{+-}\right)\label{eq:sigma_TT}\\
\frac{d\sigma_{L'T}}{dt\,dx_{B}dQ^{2}d\varphi} & =-\frac{\Gamma}{2\pi\epsilon}\,\left({\rm Im}\left(\sigma_{+0}+\sigma_{-0}\right)-\frac{1}{2}\sqrt{\frac{1-\epsilon}{1+\epsilon}}{\rm Im}\left(\sigma_{-0}-\sigma_{+0}\right)\right)\label{eq:sigma_LPrimeT}\\
\frac{d\sigma_{T'T}}{dt\,dx_{B}dQ^{2}d\varphi} & =-\frac{\Gamma}{2\pi\epsilon}\,{\rm Im}\left(\sigma_{+-}\right)\label{eq:sigma_TPrimeT}
\end{align}
and the subindices $\alpha,\beta$ in 
\[
\sigma_{\alpha\beta}=\sum_{\nu\nu'}\mathcal{A}_{\nu'0,\nu\alpha}^{*}\mathcal{A}_{\nu'0,\nu\beta},
\]
refer to polarizations of intermediate heavy boson in the amplitude
and its conjugate. As we will see below, in JLAB kinematics the contribution
of higher twist corrections is small, and for this reason, similar to the
Bethe-Heitler case, we will quantify their size in terms of the angular
harmonics $c_{n},s_{n}$, normalizing the total cross-section to the
cross-section of the dominant DVMP process as defined in~(\ref{eq:Harmonics}).

\section{Results and discussion}

\label{sec:Results}

In this section we would like to present the numerical results for the
charged current pion production . For the sake of definiteness, for
numerical estimates we use the Kroll-Goloskokov parametrization of
GPDs~\cite{Goloskokov:2006hr,Goloskokov:2007nt,Goloskokov:2008ib,Goloskokov:2009ia,Goloskokov:2011rd},
and assume the asymptotic form of the pion wave function,~$\phi_{2}(z)=6\,z\,\left(1-z\right)$.
For estimates of the twist-3 contribution introduced in Section~\ref{sec:DVMP_Xsec},
we use the parametrization suggested in~\cite{Goloskokov:2009ia,Goloskokov:2011rd},
\begin{equation}
\phi_{3}\left(z,\,l_{\perp}\right)=\phi_{3;p}\left(z,\,l_{\perp}\right)+2\phi_{3;\sigma}\left(z,\,l_{\perp}\right)=\frac{2a_{p}^{3}}{\pi^{3/2}}l_{\perp}\phi_{as}(z)\exp\left(-a_{p}^{2}l_{\perp}^{2}\right).\label{eq:phi_3}
\end{equation}
where the numerical constant $a_{p}$ is taken as $a_{p}\approx2\,{\rm GeV}^{-1}$.

\begin{figure}
\includegraphics{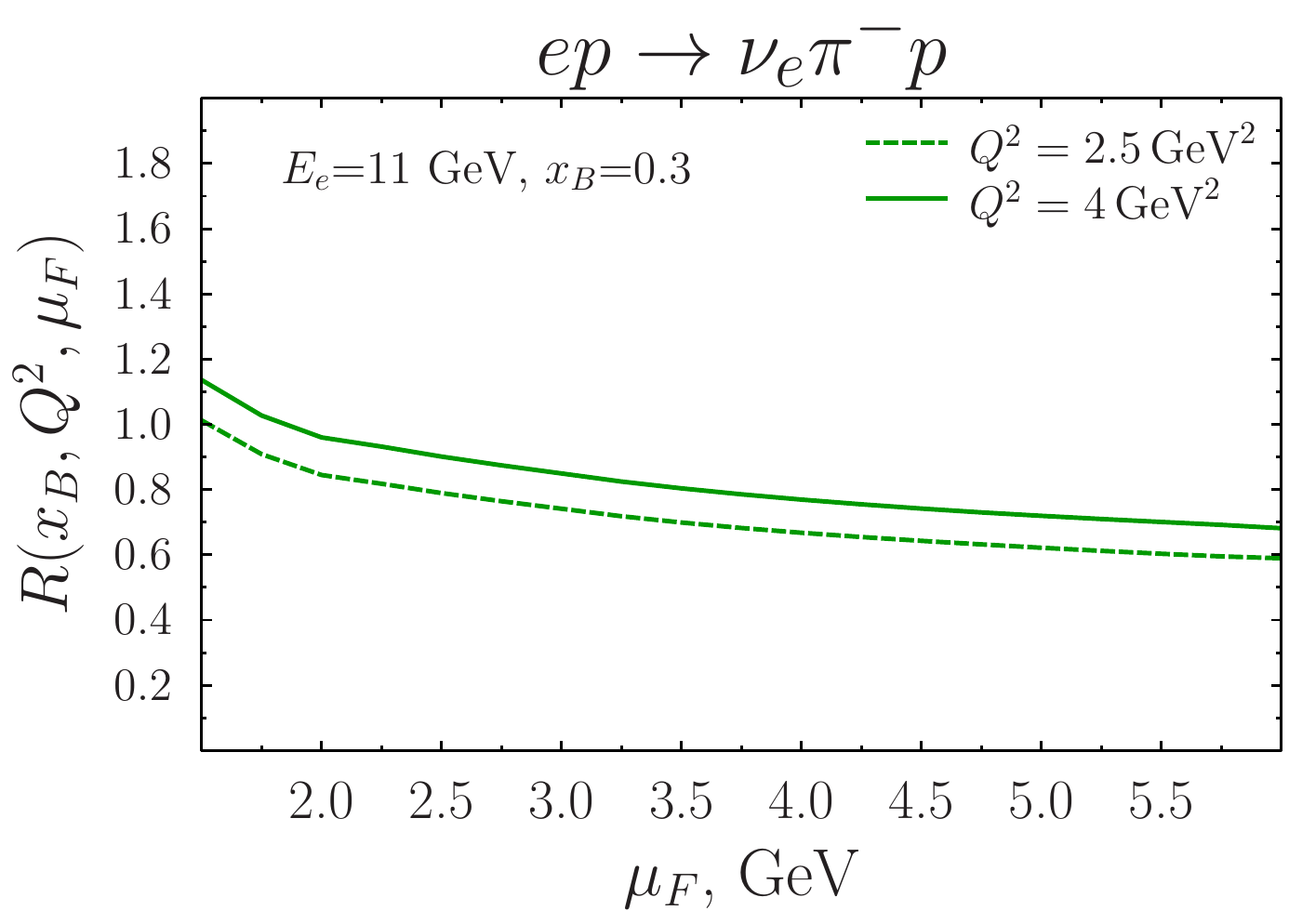}

\protect\caption{\label{fig:DVMP-pions-off}(color online) Factorization scale dependence
of the charged current process $ep\to\nu_{e}\pi^{-}p$. The ratio $R$ is defined as $R\left(x_{B},\,Q^{2},\,\mu_{F}\right)=d\sigma\left(x_{B},\,Q^{2},\,\mu_{F}\right)/d\sigma\left(x_{B},\,Q^{2},\,\mu_{F}=Q\right)$.
A similar dependence is observed for all other processes.}
\end{figure}

We would like to start with a discussion of the dependence on the factorization
scale $\mu_{F}$, which separates hard and soft physics. As we can
see from Figure~\ref{fig:DVMP-pions-off}, the dependence on the
factorization scale $\mu_{F}$ is mild and disappears for $\mu_{F}\gtrsim5$~GeV.
Though the choice of factorization scale $\mu_{F}$ is arbitrary,
taking its value significantly different from the virtuality $Q$
would lead to large logarithms in higher order corrections. As was
suggested in~\cite{Belitsky:2001nq,Ivanov:2004zv,Diehl:2007hd},
varying the scale in the range $\mu_{F}\in\left(Q/2,\,2Q\right)$,
we can roughly estimate the error due to omitted higher order loop
contributions.

\begin{figure}
\includegraphics[scale=0.6]{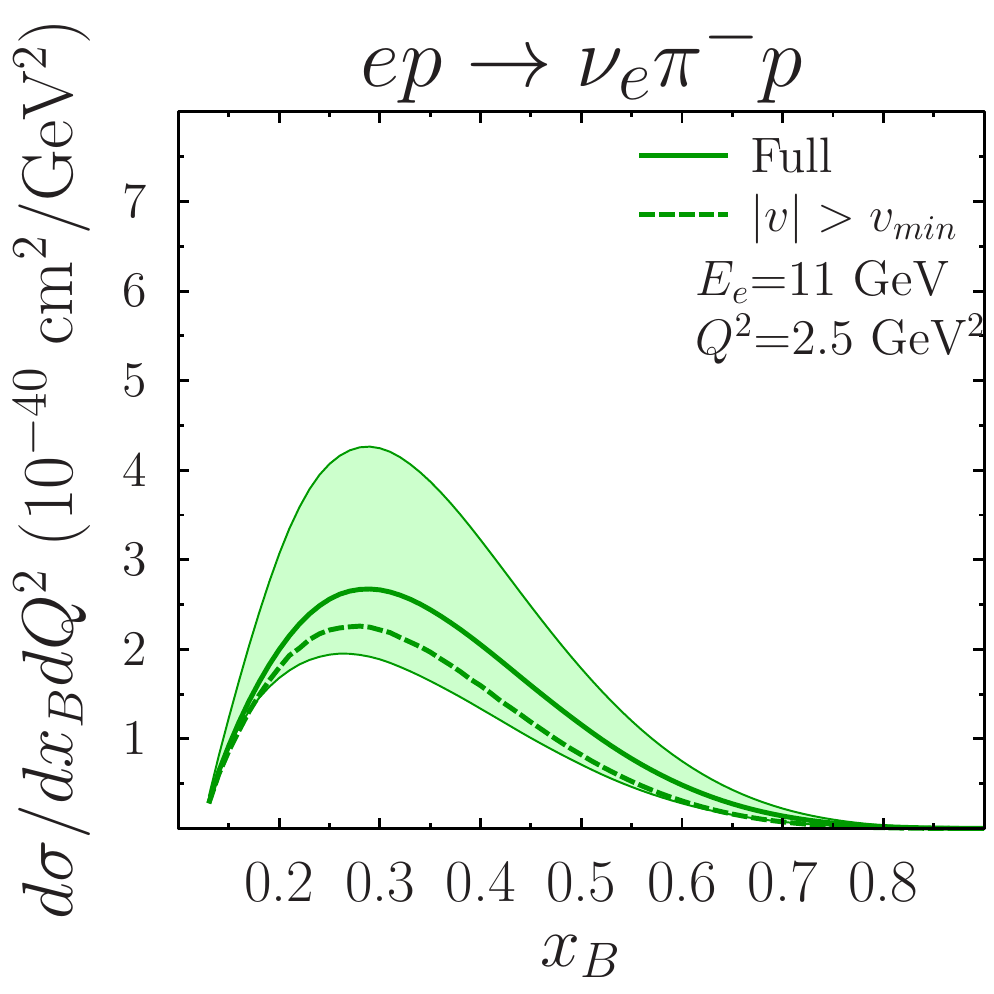}\includegraphics[scale=0.6]{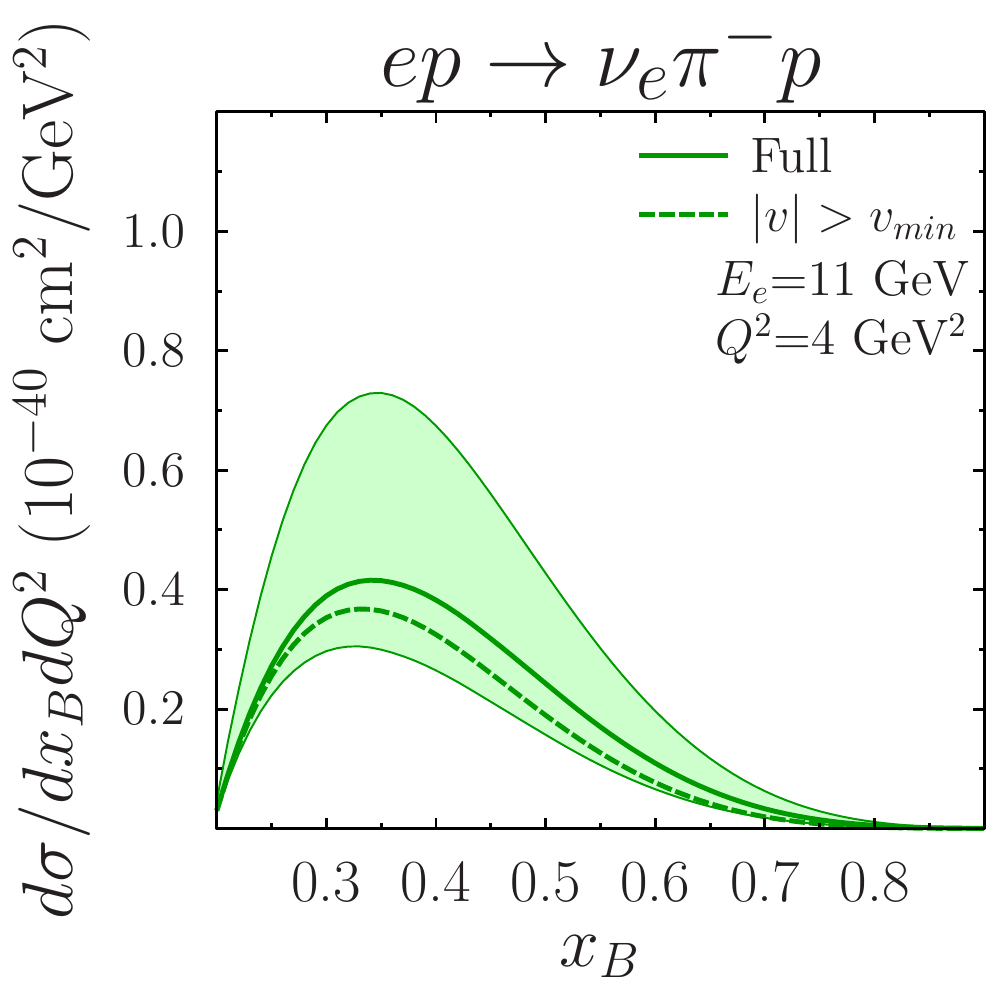}\includegraphics[scale=0.6]{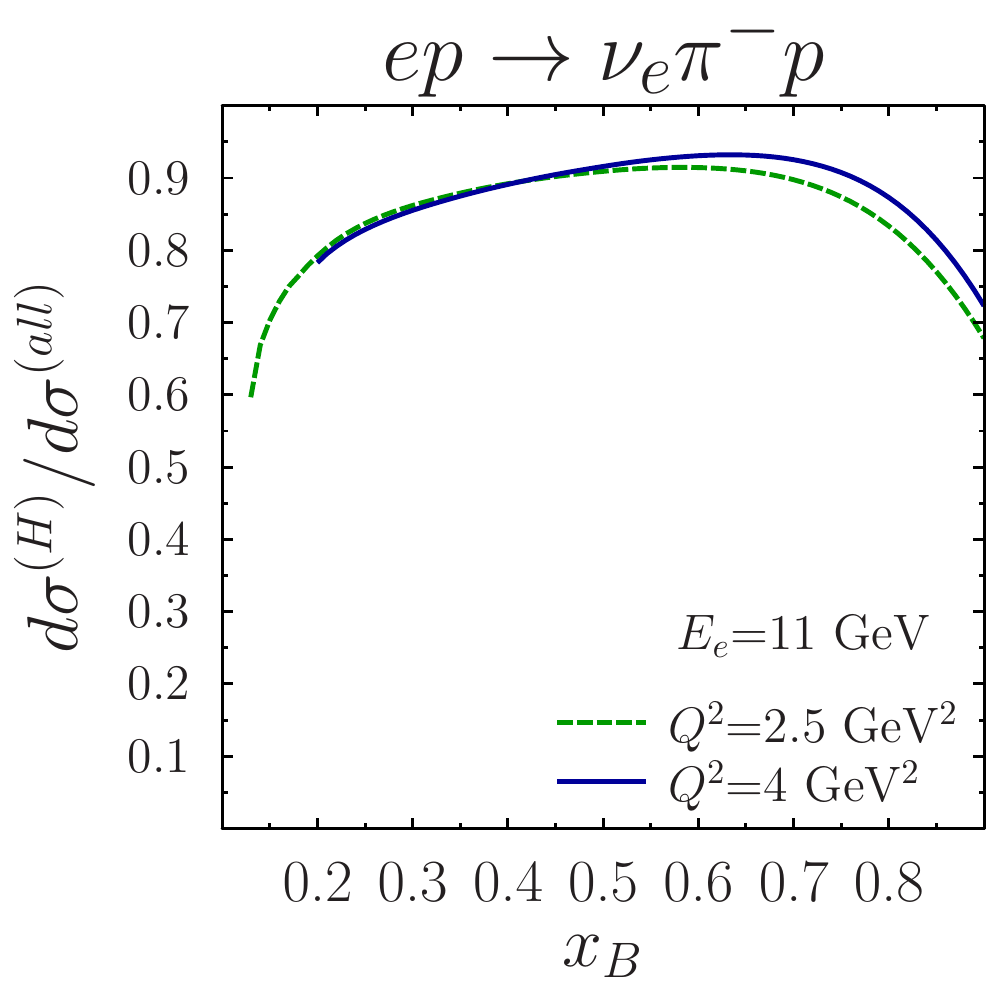}

\protect\caption{\label{fig:DVMP-pions}(color online) Left and central plots: Charged
current pion production cross-section on a proton target at fixed electron energy,
for virtualities $Q^{2}$=2.5 GeV$^{2}$(left) and $Q^{2}$=4
GeV$^{2}$(center) . Evaluations are performed using NLO coefficient
functions, as discussed in Section~\ref{sec:DVMP_Xsec}. The width
of the band represents the uncertainty due to the factorization scale
choice~$\mu_{F}\in\left(Q/2,\,2Q\right)$, as explained in the text.
The dashed line corresponds to evaluation with omitted NLO corrections
in the region of $x\approx\pm\xi$, where collinear factorization
might give sizable corrections $\sim\mathcal{O}\left(l_{\perp}^{2}/Q^{2}\right)$.
Right plot: Relative contribution of GPDs $H^{u},H^{d}$ to the total
result.}
\end{figure}

In Figure~\ref{fig:DVMP-pions} we show the predictions for the differential
cross-section $d\sigma/dx_{B}\,dQ^{2}$ for charged pion production
for two virtualities $Q^{2}$. At fixed electron energy $E_{e}$ and
virtuality $Q^{2}$, the cross-section as a function of $x_{B}$ has
a a similar bump-like shape, which is explained by an interplay of
two factors. For small $x_{B}\sim Q^{2}/2m_{N}E_{e}$ the elasticity
$y$ defined in~(\ref{eq:elasticity}) approaches one, which causes
a suppression due to a prefactor in~(\ref{eq:sigma_def}). In the
opposite limit, the suppression $\sim(1-x)^{n}$ is due to the implemented
parametrization of GPDs. Since the contribution of NLO terms is sizable,
for its evaluation we use coefficient functions which account for NLO corrections.
To estimate the uncertainty due to higher order corrections (represented
by the green band), we varied the factorization scale $\mu_{F}$ in the
range~$\mu_{F}\in\left(Q/2,\,2Q\right)$.  As was discussed in Section~\ref{sec:DVMP_Xsec},
the coefficient functions~(\ref{eq:T1},\ref{eq:T1-a},\ref{eq:T1-b})
have nonanalytic behavior $\sim\ln^{2}v$ in the region of small-$v,\bar{v}=(\xi\pm x)/2\xi$,
and therefore this region requires 
special attention. Physically, collinear factorization is not valid
here and the transverse momenta of mesons become important.
In order to assess the relative contribution of this region, we performed an
evaluation with NLO corrections switched off in the range $|v|\lesssim l_{\pi,\perp}^{2}/Q^{2}$,
where the average transverse momentum of the pion~$l_{\pi,\perp}\approx0.3-0.4\,{\rm GeV}$
was estimated from the pion charge form factor~\cite{Amendolia:1984nz,Dally:1982zk,Schlumpf:1994bc}.
As we can see from a comparison of solid and dashed lines, the contribution
of the small-$|v|$ region is quite small, and therefore we expect
that collinear factorization should give a reliable estimate in the
considered kinematics. In the rightmost pane of the Figure~\ref{fig:DVMP-pions},
we have shown the relative (dominant) contribution of the GPDs $H^{u},H^{d}$
to the total result. Contributions of helicity flip and gluon GPDs
constitute a minor ($\sim$10\%) correction to the full cross-section.

\begin{figure}
\includegraphics[scale=0.8]{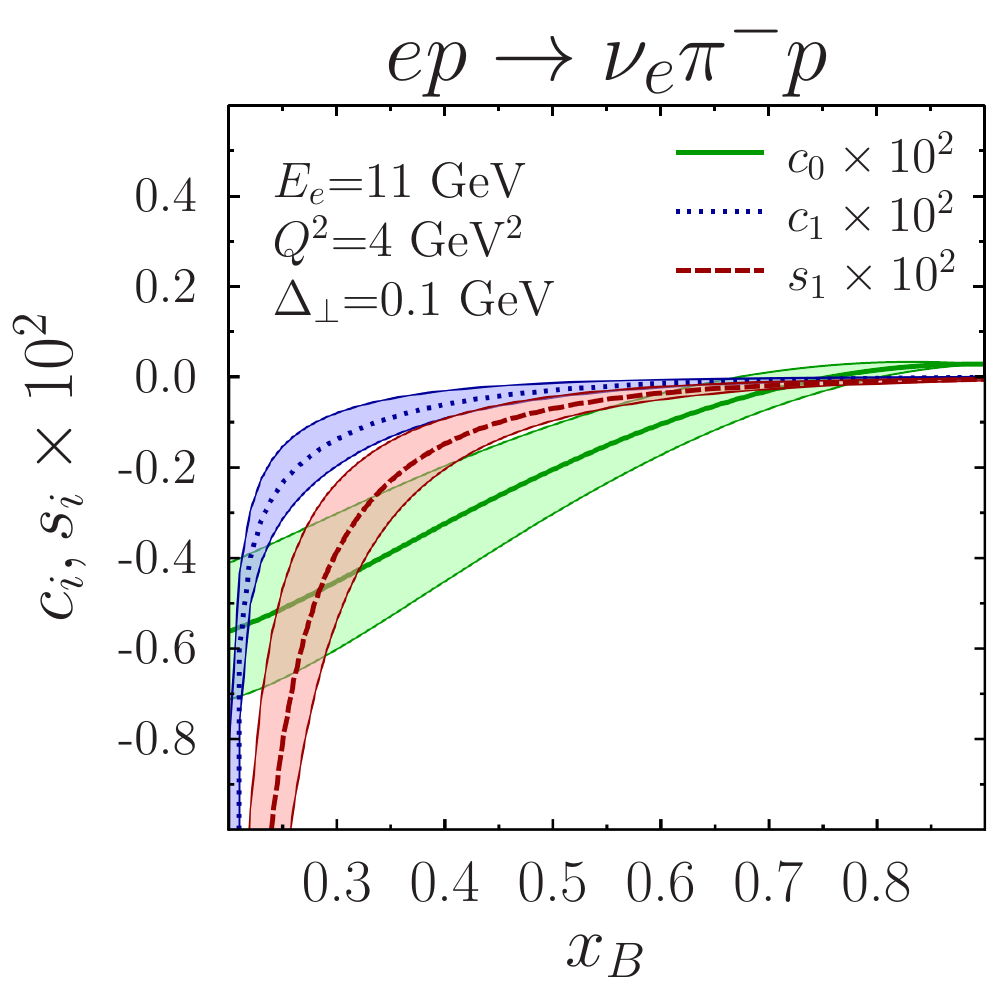}\includegraphics[scale=0.8]{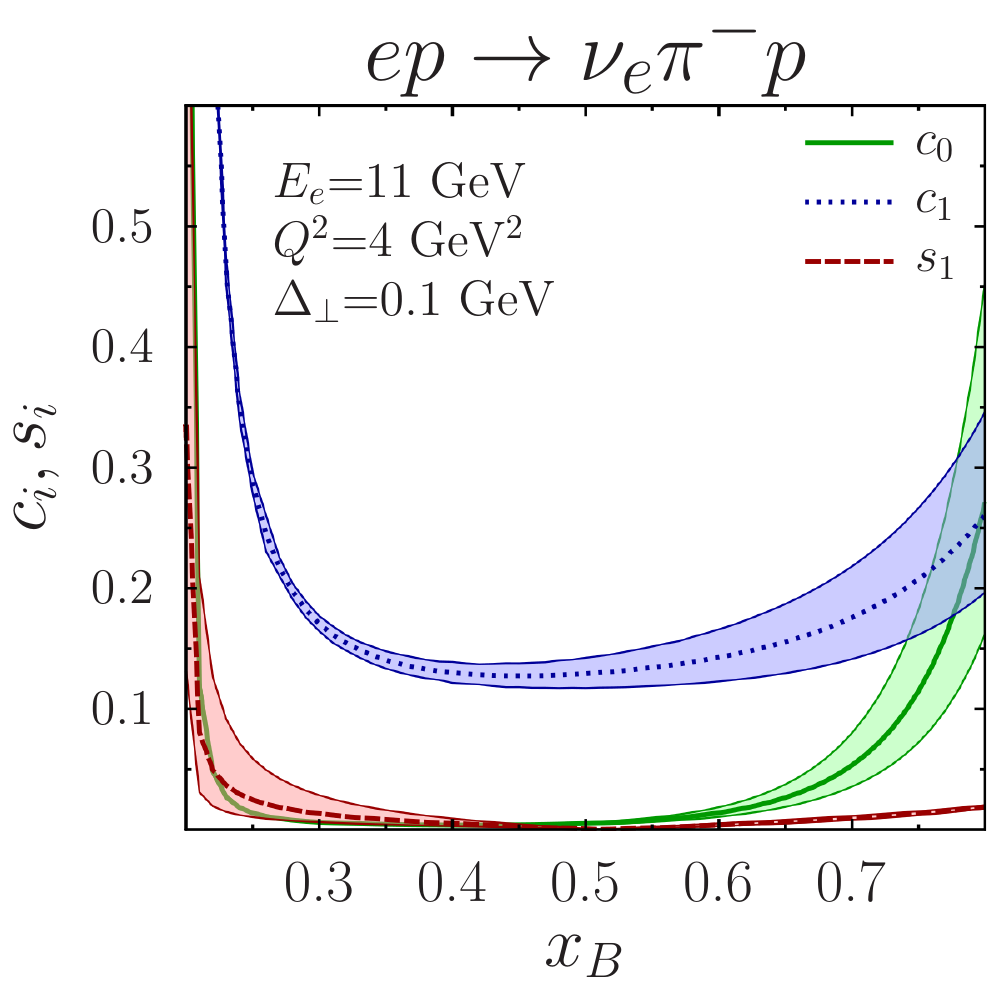}

\protect\caption{\label{fig:DVMP-pions-Asy}(color online) Left: Harmonics $c_{n},\,s_{n}$
in charged current pion production on a proton target, due to the interference
with the Bethe-Heitler contribution. Right: Harmonics $c_{n},\,s_{n}$
generated due to twist-2 and twist-3 GPDs interference. In
both plots the evaluations are performed using NLO coefficient functions,
as discussed in Section~\ref{sec:DVMP_Xsec}. The width of the band
represents the uncertainty due to the factorization scale choice~$\mu_{F}\in\left(Q/2,\,2Q\right)$,
as explained in the text.}
\end{figure}

The contribution of the asymptotic Bethe-Heitler mechanism introduced
in Section~\ref{sec:BH} is shown in the left pane of the Figure~\ref{fig:DVMP-pions-Asy}.
We can see that for 11 GeV electron beams, its contribution is small
and does not exceed $\sim1\,$ per cent. The smallness of the harmonics
$c_{n},\,s_{n}$ is explained by the fact that the kinematic
prefactor $\sim\left(Q^{2}/t\right)$ enhancement  in JLAB kinematics is not sufficient
to compensate the suppression $\sim\mathcal{O}\left(\alpha_{{\rm em}}\right)$.
Though formally both the BH term~(\ref{eq:XSec_BH}) and interference
term~(\ref{eq:XSec_Int}) lead to appearance of the harmonics $c_{0},...,c_{2}$,
the contribution from the former is suppressed by an additional
power of $\alpha_{{\rm em}}$ and thus in JLAB kinematics the harmonics get a
major contribution from the interference term. For the same reason, the
harmonics $c_{2}$ (not shown in the plot) is extremely small: it
gets contribution only from BH. In the right pane of the plot we have
shown similar harmonics generated due to twist-three interference.
The largest harmonics $c_{1}$ does not exceed 20 per cent and after
averaging over the angle $\varphi$ does not contribute to the total
cross-section $d\sigma/dx_{B}dQ^{2}$. The harmonics $c_{0}$, which
contributes to the integrated cross-section $d\sigma/dx_{B}dQ^{2}$
as a multiplicative factor $1+c_{0}$, in the region of interest ($x_{B}\approx0.4\pm0.2$)
is small and constitutes a few percent correction.

\begin{figure}
\includegraphics[scale=0.6]{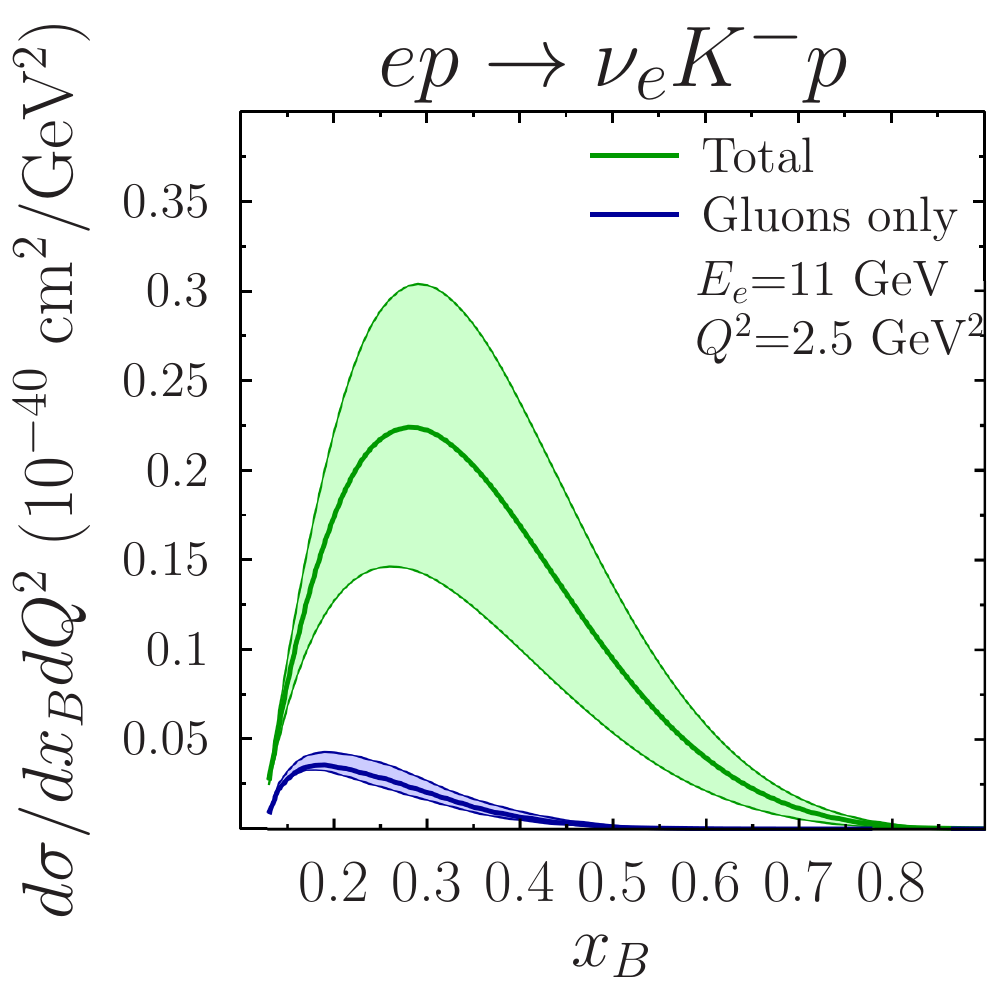}\includegraphics[scale=0.6]{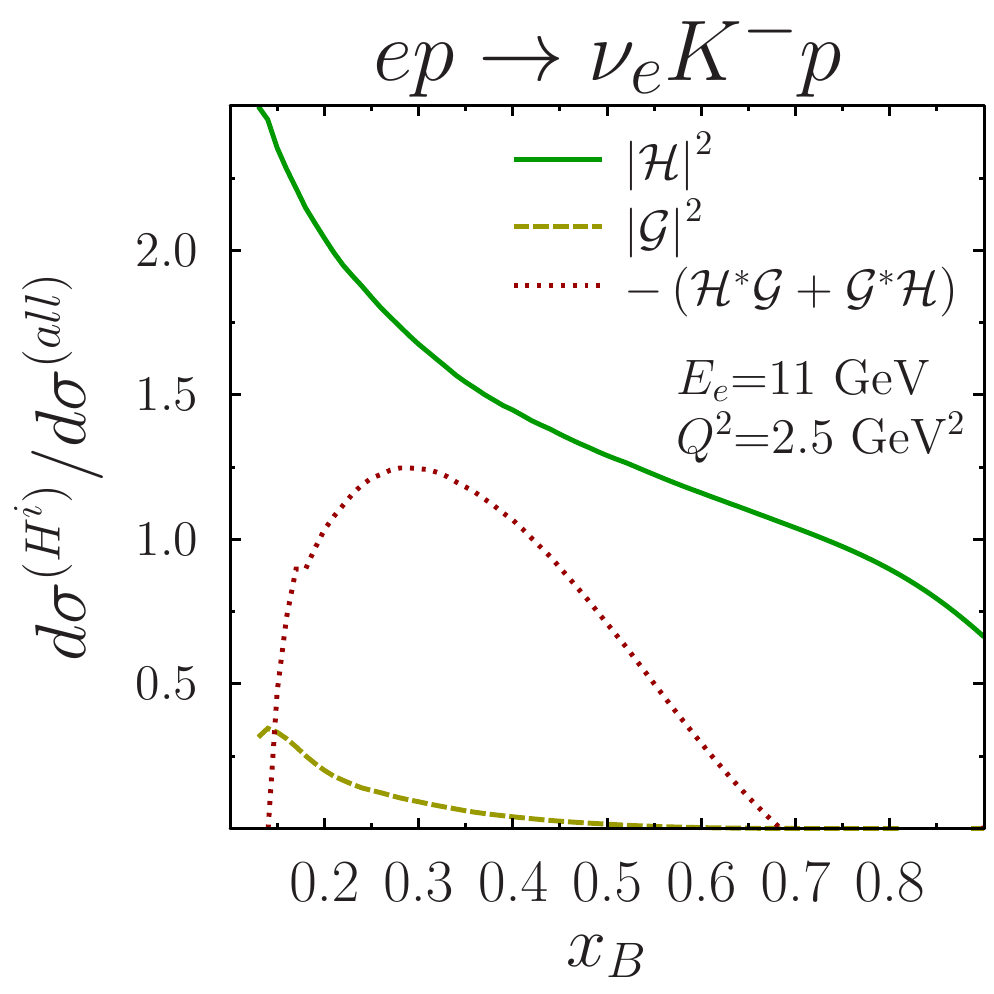}\includegraphics[scale=0.6]{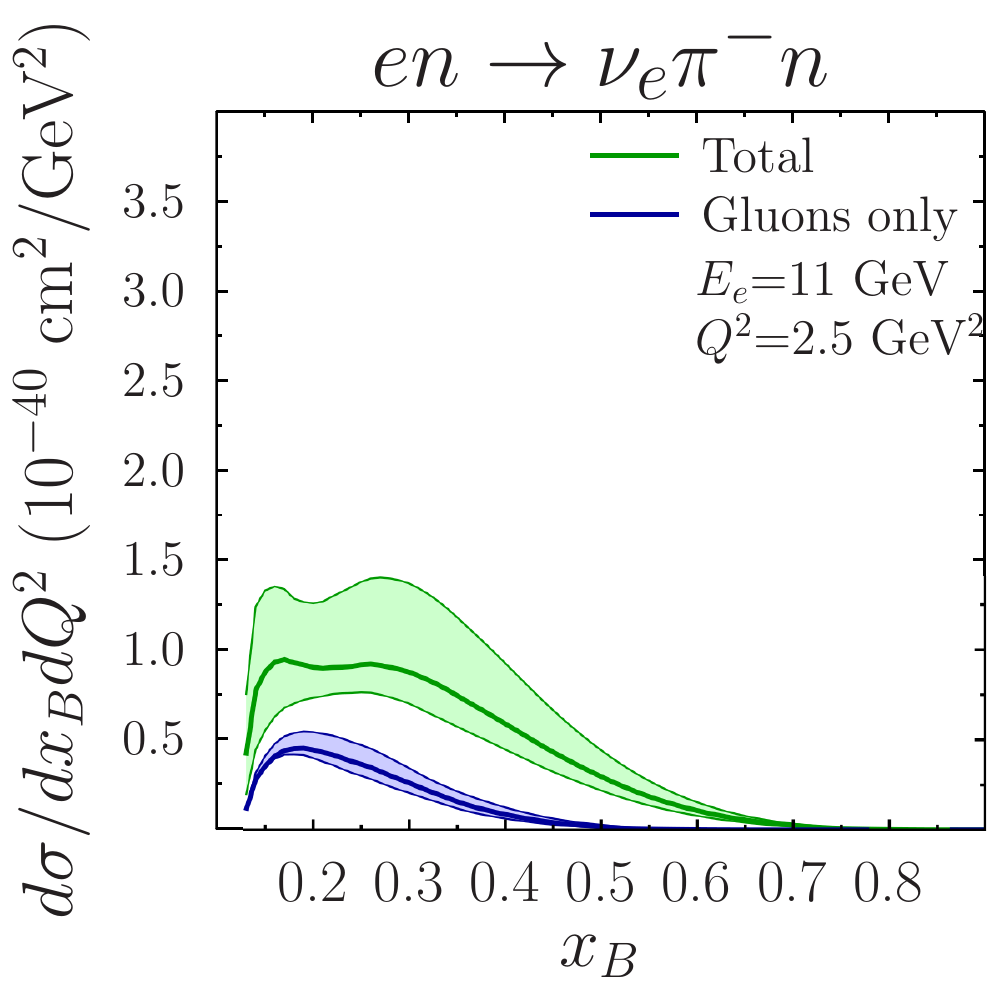}

\protect\caption{\label{fig:DVMP-kaons}(color online) Left: Charged current kaon production
cross-sections for fixed energy electron beam ($E_{e}\approx11$~GeV).
Central: relative fractions of different GPD components to the kaon
production cross-section. We can see that the interference between
quark and gluons is large and contributes with a negative sign. Right:
Charged current pion production on a neutron target. In all plots the
evaluations are performed using NLO coefficient functions, as discussed
in Section~\ref{sec:DVMP_Xsec}. The width of the band represents
the uncertainty due to the factorization scale choice~$\mu_{F}\in\left(Q/2,\,2Q\right)$,
as explained in the text.}
\end{figure}

For deeply virtual meson production in other channels the cross-section
gets comparable contributions from GPDs of different partons. For
this reason restrictions imposed by experimental data on GPDs of individual
partons are less binding. Additionally, these channels present more
challenges for experimental study. For example, for charged current
kaon production (see left pane in the Figure~\ref{fig:DVMP-kaons}),
we observe that the cross-section is small due to Cabibbo suppression
($\Delta S=1$), so the statistical error will be larger. From the
central pane in the Figure~\ref{fig:DVMP-kaons} we can see that
the total cross-section of this process gets a sizable contribution
from quark-gluon interference. Similarly, for charged current
pion production on a neutron (right pane in the Figure~\ref{fig:DVMP-kaons}),
the cross-section gets significant contributions from gluon GPDs and
its interference with quarks, and experimentally the precision will
be affected by uncertainty in the reconstruction of scattered neutron
kinematics.

For this reason we believe that the study of the GPDs with charged
currents should be focused on the $ep\to\nu_{e}\pi^{-}p$ channel.

\section{Conclusions}

In this paper we have shown that generalized parton distributions can
be probed in charged current meson production processes, $ep\to\nu_{e}\pi^{-}p$.
In contrast to pion \emph{photo}production, these processes get a major
contribution from the unpolarized GPDs $H^{u},\,H^{d}$, and thus
could be used to supplement studies of these GPDs in DVCS. The undetectability
of the produced neutrino will not present major challenges for the
kinematics reconstruction, since all final state hadrons are charged. We estimated
the cross-sections in the kinematics of the upgraded 12 GeV Jefferson
Laboratory experiments and found that thanks to the large luminosity, the
process can be measured with reasonable statistics. We also estimated the
contaminating contributions from the Bethe-Heitler mechanism and twist-three
corrections due to transversity GPDs. We found that both are small, and
for this reason the $ep\to\nu_{e}\pi^{-}p$ channel presents a clean probe
of the target GPDs . If polarized targets become available
in these experiments, it could enable to study various beam-target asymmetries,
sensitive to the smaller GPDs $E,\,\tilde{H},\,\tilde{E}$.

A code for the evaluation of the cross-sections, with various GPD models,
is available on demand.

\section*{Acknowledgments}

This research was partially supported by Proyecto Basal FB 0821 (Chile),
the Fondecyt (Chile) grants 1140390 and 1140377, CONICYT (Chile) grant
PIA ACT1413. Powered@NLHPC: This research was partially supported
by the supercomputing infrastructure of the NLHPC (ECM-02). Also,
we thank Yuri Ivanov for technical support of the USM HPC cluster
where part of evaluations were done.

\appendix

 \end{document}